\documentclass[fleqn,usenatbib]{mnras}
\pdfoutput=1

\usepackage{newtxtext,newtxmath}

\usepackage{hyperref}
\usepackage{cleveref}

\usepackage[T1]{fontenc}
\usepackage{ae,aecompl}

\usepackage{graphicx}	
\usepackage{amsmath}	
\usepackage{amssymb}	

\usepackage{color}
\usepackage{xspace}
\usepackage{enumitem}

\usepackage{booktabs}
\usepackage{array}
\newcolumntype{L}[1]{>{\raggedright\let\newline\\\arraybackslash\hspace{0pt}}m{#1}}
\newcolumntype{C}[1]{>{\centering\let\newline\\\arraybackslash\hspace{0pt}}m{#1}}
\newcommand{\ra}[1]{\renewcommand{\arraystretch}{#1}}

\graphicspath{.}

\newcommand{\rtt}{\mathrm{R}_{23}}
\newcommand{\ott}{\mathrm{O}_{32}}

\newcommand{\llambda}{\ensuremath{\lambda \lambda}}
\newcommand{\SN}{\mathrm{S/N}}

\newcommand{\Ha}{{{\rm H}\ensuremath{\alpha}}}
\newcommand{\Hb}{{{\rm H}\ensuremath{\beta}}}
\newcommand{\OII}{{\rm [\ion{O}{II}]}}
\newcommand{\OIII}{{\rm [\ion{O}{III}]}}
\newcommand{\NII}{{\rm [\ion{N}{II}]}}
\newcommand{\SII}{{\rm [\ion{S}{II}]}}
\newcommand{\HII}{\ion{H}{II}}
\newcommand{\OH}{{\rm O}/{\rm H}}

\newcommand{\dex}{\,{\rm dex}}

\newcommand{\cms}{\,\mathrm{cm}\,\mathrm{s}^{-1}}

\newcommand{\Msun}{\,{\rm M}_{\odot}}
\newcommand{\mum}{\,{\rm \ensuremath{\mu}m}}
\newcommand{\peryr}{\,{\rm yr}^{-1}}

\newcommand{\Msunyr}{\ensuremath{\,\Msun\, {\rm yr}^{-1}}}
\newcommand{\ang}{\ensuremath{\textrm{\AA}}\xspace}
\newcommand{\met}{12+\log(\mathrm{O/H})}

\newcommand{\Mstar}{{\rm{M}\ensuremath{_{*}}}}
\newcommand{\SFR}{{\rm SFR}}
\newcommand{\sSFR}{{\rm sSFR}}

\newcommand{\cmcub}{\ensuremath{\,{\rm cm}^{-3} }}

\newcommand{\kmsMpc}{\ensuremath{ \, {  \rm km \, s^{-1} \, Mpc^{-1} }   } }

\title[The ionisation parameter of star-forming galaxies evolves with the sSFR]{The ionisation parameter of star-forming galaxies evolves with the specific star formation rate}

\author[M. Kaasinen et al.]{
Melanie Kaasinen$^{1,2,3}$\thanks{E-mail: kaasinen@mpia.de},
Lisa Kewley$^{1,4}$,
Fuyan Bian$^{1,5}$,
Brent Groves$^{1,4}$, \newauthor
Daichi Kashino$^{6}$, John Silverman$^{7}$, Jeyhan Kartaltepe$^{8}$     
\\
$^{1}$Research School of Astronomy and Astrophysics, Australian National University, Weston Creek 2611, Australia \\
$^{2}$Max-Planck-Institut f\"ur Astronomie, K\"onigstuhl 17, D-69117 Heidelberg, Germany \\
$^{3}$Universit\"at Heidelberg, Zentrum f\"ur Astronomie, Institut f\"ur Theoretische Astrophysik, Albert-Ueberle-Stra{\ss}e 2, 69120 \\ Heidelberg, Germany \\
$^{4}$ARC Centre of Excellence for All Sky Astrophysics in 3 Dimensions (ASTRO 3D) \\
$^{5}$Stromlo Fellow\\
$^{6}$Department of Physics, ETH Z\"urich, Wolfgang-Pauli-Strasse 27, CH-8093 Z\"urich, Switzerland\\
$^{7}$Kavli Institute for the Physics and Mathematics of the Universe, The University of Tokyo, Kashiwa, Japan 277-8583 \\ (Kavli IPMU, WPI)\\
$^{8}$School of Physics and Astronomy, Rochester Institute of Technology, 84 Lomb Memorial Drive, Rochester NY 14623, USA 
}

\date{Accepted 19 April 2018. Received YYY; in original form ZZZ}

\pubyear{2018}

\begin{document}
\label{firstpage}
\pagerange{\pageref{firstpage}--\pageref{lastpage}}
\maketitle

\begin{abstract} 
    We investigate the evolution of the ionisation parameter of star-forming galaxies using a high-redshift ($z\sim 1.5$) sample from the FMOS-COSMOS survey and matched low-redshift samples from the Sloan Digital Sky Survey. By constructing samples of low-redshift galaxies for which the stellar mass (\Mstar), star formation rate (SFR) and specific star formation rate (sSFR) are matched to the high-redshift sample we remove the effects of an evolution in these properties. We also account for the effect of metallicity by jointly constraining the metallicity and ionisation parameter of each sample.   We find an evolution in the ionisation parameter for main-sequence, star-forming galaxies and show that this evolution is driven by the evolution of sSFR. By analysing the matched samples as well as a larger sample of $z<0.3$, star-forming galaxies we show that high ionisation parameters are directly linked to high sSFRs and are not simply the byproduct of an evolution in metallicity. Our results are physically consistent with the definition of the ionisation parameter, a measure of the hydrogen ionising photon flux relative to the number density of hydrogen atoms. 
\end{abstract}



\section{Introduction} 

    Most high-redshift ($z>1$) star-forming galaxies appear to have significantly different properties to the dominant population of local star-forming galaxies. High-redshift (high-z) star-forming galaxies typically have 10-100 times higher star formation rates (SFRs) \citep{2011A&amp;A...533A.119E,2007ApJ...670..156D,2014ApJS..214...15S},  $3-10$ times higher gas fractions \citep{2008ApJ...687...59G,2010Natur.463..781T} and higher gas velocity dispersions \citep{2006ApJ...646..107E,2006ApJ...645.1062F,2008ApJ...687...59G,2007ApJ...670..156D,2010Natur.463..781T} than local star-forming galaxies of the same stellar mass (\Mstar). High-z galaxies also appear to be smaller by $0.3-0.5 \dex$ \citep{2007ASSP....3..481T,2014ApJ...788...28V} and more ``clumpy'' \citep[e.g.][]{2005ApJ...627..632E} than local star-forming galaxies. 

    The differences between local and high-z star-forming galaxies are largely driven by the conditions within the inter-stellar medium (ISM). The most convenient way to probe the ISM conditions of both local and high-redshift star-forming galaxies is to study their rest-frame optical emission. This emission is typically dominated by a small set of strong emission lines, consisting of hydrogen recombination lines and collisionally excited metal lines, which stem from regions of the ISM that have been ionised by recent ($<5\, \mathrm{Myr}$) star formation. The relative strengths of these strong emission lines reflect the ionising sources and physical conditions of the ionised ISM, particularly of the most luminous \HII\ regions.  

    Over the last decade, large samples of high-redshift, star-forming galaxies with rest-frame optical emission-line measurements have been assembled \citep[e.g.][]{2014ApJ...785..153M,0004-637X-801-2-88,2015ApJS..220...12S}. The ensueing studies have found that high redshift star-forming galaxies exhibit emission-line ratios that are atypical of the local star-forming galaxy population \citep[e.g.][]{0004-637X-701-1-52,2013ApJ...774L..10K,2014ApJ...795..165S,2015PASJ...67...80H,0004-637X-801-2-88}. In particular, there is increasing evidence for an enhancement of the $\OIII \llambda 4959,5007/\OII \llambda 3726,3729$ and $\OIII\lambda 5007/\Hb$ ratios with respect to local galaxies. These elevated emission-line ratios indicate that at least some ISM conditions must have evolved since the early Universe. But it is still unclear which ISM conditions are evolving and to what extent.

    The conditions within the ionised ISM can be described by a small set of physical properties including the metallicity, ionisation parameter, pressure and hardness of the ionising radiation field \citep[e.g.][]{0004-637X-647-1-244,0004-637X-774-2-100,0067-0049-208-1-10,2014ApJ...793..127V}. The gas-phase metallicity of typical, high-z, star-forming galaxies is lower than in the local population \citep[e.g.][]{2006ApJ...646..107E,2008ApJ...678..758L,2013ApJ...771L..19Z}. However, the evolution of metallicity alone is not sufficient to account for the elevated $\OIII \llambda 4959,5007/\OII \llambda 3726,3729$ and $\OIII\lambda 5007/\Hb$ ratios at high redshift \citep[e.g.][]{2014ApJ...795..165S,2014ApJ...785..153M,2013ApJ...774L..10K,2014ApJ...787..120S}. Other ISM conditions that may account for the observed changes include higher ionisation parameters and/or electron densities \citep[e.g.][]{2008MNRAS.385..769B,0004-637X-774-2-100,2014ApJ...787..120S,2014ApJ...795..165S,2015ApJ...812L..20K}, harder ionising radiation fields \citep{0004-637X-774-2-100,2014ApJ...795..165S}, contributions from shocks/AGN \citep{2006MNRAS.371.1559G,2014ApJ...781...21N} and a variation in N/O ratio \citep[e.g.][]{0004-637X-801-2-88,2016ApJ...828...18M}. 

    Many studies find evidence for increased ionisation parameters at high redshift \citep[e.g.][]{0004-637X-701-1-52,2010ApJ...725.1877B,2014MNRAS.440.2201S,2014ApJ...787..120S,2014ApJ...785..153M,2016ApJ...816...23S}.  The ionisation parameter, $q$, is a measure of the current star formation distribution and ionisation state of the ISM, reflecting the interaction between the ionising source and ionised gas. Defined as the ratio between the mean hydrogen ionising photon flux and the density of hydrogen atoms \citep[e.g][]{2003adu..book.....D,2006agna.book.....O}, the ionisation parameter increases with the luminosity of the stellar population and the hardness of the ionising radiation field. 

    Although there are a range of physical mechanisms that can increase the ionisation parameter of \ion{H}{II} regions it is still unclear what is driving the high ionisation parameters of high-z, star-forming galaxies. High-redshift studies have often been limited by small sample sizes with insufficient emission lines that are sensitive to the ionisation parameter, relying on the use of a single emission-line ratio as a proxy for the ionisation parameter. \cite{2014MNRAS.442..900N} suggest that the high inferred ionisation parameters at high redshift are the byproduct of lower metallicities, but the inverse correlation between the ionisation parameter and metallicity of \ion{H}{II} regions \citep{0004-637X-647-1-244} has not yet been demonstrated on a global scale. It has also been suggested that high ionisation parameters are the result of high SFRs, which lead to a larger reservoir of ionising photons \citep[e.g.][]{0004-637X-701-1-52,0004-637X-774-2-100}. Although many of the galaxies with high ionisation parameters have high SFRs, both locally and at high redshift, there is still no direct evidence for an increase in ionisation parameter with SFR. 

    Previous observational studies have struggled to disentangle the evolution of the ionisation parameter from both selection effects and the evolution of global galaxy properties. High-redshift studies typically sample galaxies with intrinsically luminous emission lines \citep{2014ApJ...796..102C,2014ApJ...788...88J}, such as starburst galaxies or galaxies containing significant contributions from shocks or AGN \citep{2006MNRAS.371.1559G,2011ApJ...743..144T}, which have higher ionisation parameters. Moreover, most high-redshift spectroscopic surveys are targeted at the dominant population of star-forming galaxies, i.e. main-sequence galaxies (at $10^9-10^{11}\Msun$), which are offset to higher SFRs and sSFRs at high redshift \citep{2014ApJS..214...15S,2014ARA&amp;A..52..415M}. Although previous studies have provided evidence for a correlation between global galaxy properties (i.e. \Mstar\ and SFR) and the gas-phase metallicity \citep[e.g][]{2004ApJ...613..898T,2010MNRAS.408.2115M,2014ApJ...791..130Z}, the relationship with ionisation parameter remains unclear.

    Recent studies indicate that the elevated $\OIII\llambda 4959,5007/\OII 3726,3729$ and $\OIII\lambda 5007/\Hb$ ratios and high inferred values of electron density and ionisation parameter are related to higher specific star-formation rates (sSFR; SFR/\Mstar) \citep{2015ApJ...812L..20K,2016ApJ...820...73H,2016ApJ...822...62B}. \cite{2016ApJ...822...62B} find that local galaxies with the elevated \OIII/\Hb\ ratios typical of galaxies at $z\sim 2$ have significantly higher electron densities, ionisation parameters and sSFRs. Conversely, \cite{2016ApJ...828L..11D} show that high-redshift galaxies with low sSFRs ($\log(\sSFR/\peryr)<-9$) also exhibit low \OIII/\Hb\ ratios ($\log(\OIII/\Hb)<0.3$).  

    Although these studies go some way to linking global properties with ISM conditions, they do not isolate the impact that global properties such as the SFR and sSFR have on the ionisation parameter. We investigate the extent to which the \Mstar, SFR and sSFR impact the evolution of the ionisation parameter by comparing samples of low- and high-z galaxies matched in these properties. By matching samples we remove the effects of the evolution of \Mstar, SFR or sSFR with redshift. We account for the evolution of metallicity and the degeneracy between strong emission-line ratios and different ISM properties by applying diagnostic methods which simultaneously infer the metallicity and ionisation parameter.
 
    This paper is structured as follows. In Section \ref{sec:sample_selection}, we describe the main high-z sample and the matched low-z comparison samples. We describe the observations, data reduction and relative aperture correction factors for our main high-z ($z\sim 1.5$) sample in Section \ref{sec:highz_data}. In Section \ref{sec:stacking} we describe the methods used to derive stacked emission-line fluxes for our samples. We outline the methods used to diagnose the ionisation parameter and metallicity in Section \ref{sec:diagnosing}. We present our results, including the emission-line properties, metallicities and ionisation parameters in Section \ref{sec:results}. In Section \ref{sec:discussion} we discuss the validity of our methods and the relationships between the ionisation parameter, metallicity and global galaxy properties. Finally, in Section \ref{sec:conclusion}, we summarise our findings. 

    Throughout this paper we refer to values of \SFR, \sSFR\  and \Mstar\ consistent with a Kroupa IMF. We select the Kroupa IMF for consistency with the SFRs of the SDSS sample. We adopt a $\mathrm{\Lambda}$-CDM cosmology with $H_0=70\kmsMpc$, $\Omega_m = 0.3$, and $\Omega_\Lambda = 0.7$.  Throughout the paper we use ``metallicity'' and $Z$ to mean the gas-phase oxygen abundance relative to hydrogen, $\met$.

\section{Sample Selection} 
    \label{sec:sample_selection}

    \subsection{High-z sample} 
        \label{sub:high_redshift_sample}

        To investigate the evolution of the ionisation parameter, we assemble a sample of star-forming galaxies at $z\sim 1.5$ with rest-frame optical emission-line flux measurements of $\OII\llambda 3726,3729$, \Hb, $\OIII\lambda5007$, \Ha\ and $\NII\lambda6584$. Our sample is derived from two spectroscopic surveys of star-forming galaxies in the COSMic evOlution Survey (COSMOS) field. The initial survey, undertaken with the Fibre Multi-Object Spectrograph (FMOS) on Subaru (PIs Sanders and Silverman, \citealt{2015ApJS..220...12S}), yielded detections of \Hb, $\OIII\lambda5007$, \Ha\ and $\NII\lambda6584$ whereas the complementary survey undertaken with the DEep Imaging Multi-Object Spectrograph (DEIMOS) on Keck II (PI L.J. Kewley, \citealt{2017MNRAS.465.3220K}) provided observations of the $\OII$ doublet. We investigated the evolution of the electron density using the COSMOS-\OII\ sample in \citep{2017MNRAS.465.3220K}, showing that the electron density does not evolve with redshift when the changing SFR is taken into account. In this work, we investigate the evolution of the ionisation parameter and its dependence on \Mstar, SFR and sSFR.

        Both spectroscopic surveys were targeted at massive ($ > 10^{9.8}\Msun$), star-forming galaxies at $1.4<z<1.8$, identified from the COSMOS photometric catalogues \citep{2012A&amp;A...544A.156M,2013A&amp;A...556A..55I}. Initial stellar masses and photometric redshifts for target selection were estimated via the broad-band photometry and fitting results of \texttt{LePHARE} \citep{2011ascl.soft08009A}, using \cite{2003MNRAS.344.1000B} population synthesis models and a Chabrier IMF. FMOS-COSMOS targets were selected to have a photometric redshift within the optimal range for the H-long or H-short grating ($R \sim 3000$; 1.6-1.8\mum, 1.4-1.6\mum) and a \Ha\ flux above the required threshold for detectability \citep[see ][for details]{2015ApJS..220...12S}. Galaxies with \Ha\ detections were re-observed using the J-long grating ($R\sim 2200$; 1.11-1.35\mum) to detect \Hb\ and $\OIII \lambda 5007$. Follow-up J-long observations were prioritised based on the reliability of the spectroscopic redshifts. The complemetary COSMOS-\OII\ survey targeted $\sim 800$ galaxies with $\SFR_\mathrm{phot} > 10\Msunyr $ and z(AB) magnitudes $\lesssim 24$ (SuprimeCam, $z^{++}, \lambda_c = 9106$, \citealt{2016ApJS..224...24L}). To minimise AGN contamination, galaxies with X-ray detections were excluded from the COSMOS-\OII\ Survey. 

        To reliably probe the relative strengths of the five strong emission lines used in this study, $\OII\lambda3727, \Hb, \OIII\lambda5007, \Ha$ and $\NII\lambda6584$, we require observations in all three wavelength regimes. Of the $\sim 380$ galaxies with FMOS H- and J-band observations, 97 have corresponding DEIMOS observations. We discard 12 galaxies with DEIMOS spectra that do not cover the predicted wavelength of the $\OII\lambda 3726, 3729$ doublet or for which the spectrum is impacted by issues such as sky continuum errors caused by scattered light from a neighbouring slit \citep[see ][]{2013ApJS..208....5N}. Of the remaining set we remove three galaxies for which the \OIII/\Hb\ and \NII/\Ha\ ratios indicate the presence of an AGN based on the maximum starburst criteria of \cite{2001ApJ...556..121K}. Measurements of all five SELs used in this study are only available for 13 of the remaining 82 galaxies. 

\begin{figure}
          \begin{center}  
                \includegraphics[width=0.5\textwidth,trim={0.0cm 0.4cm 0.4cm 0.5cm},clip]{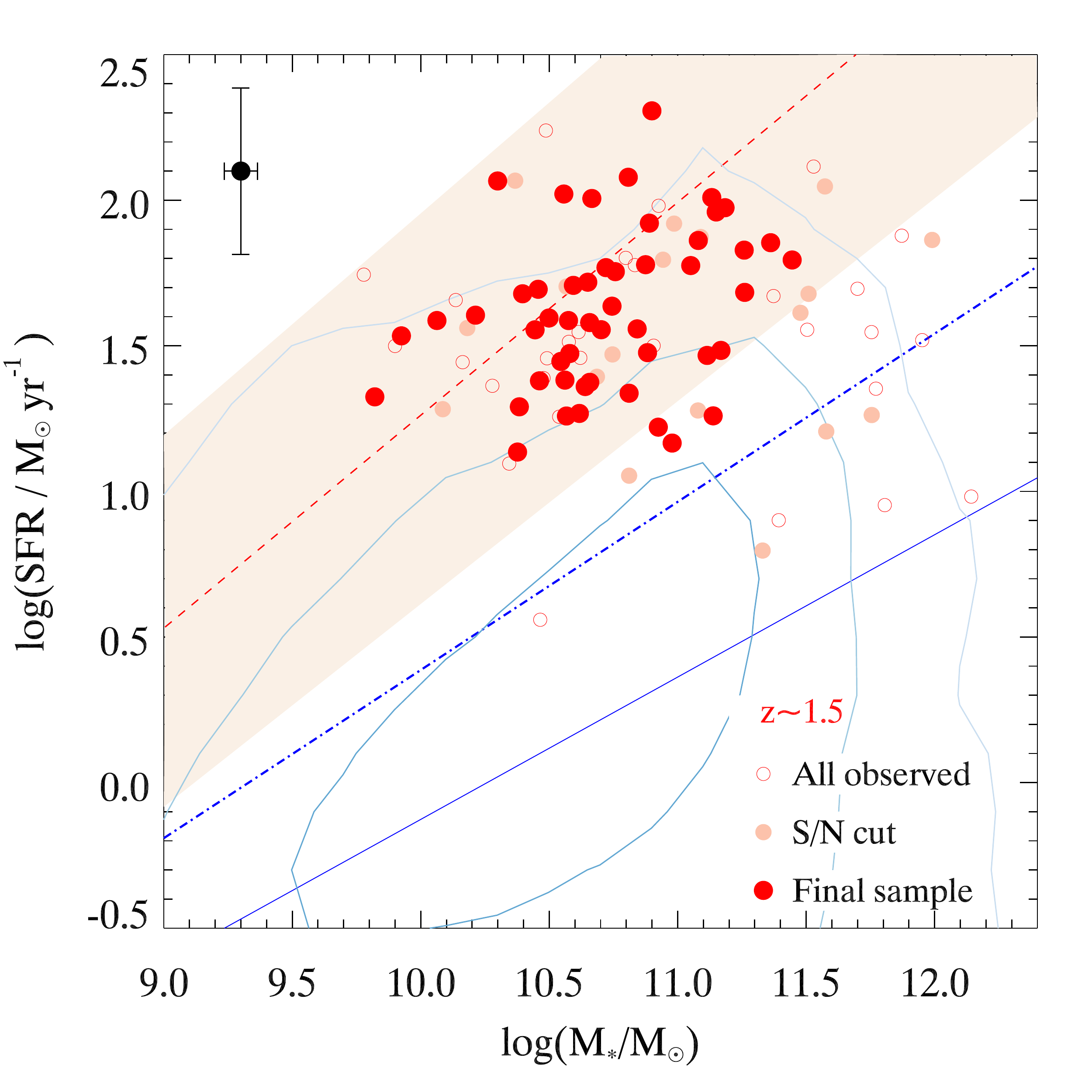} 
              \end{center}
            \caption{SFR vs \Mstar for the ``SDSS star-forming catalogue'' (blue) and the $z\sim 1.5$ main and parent samples. Open red circles denote the initial sample of $z\sim 1.5$ galaxies with FMOS H- and J-band and DEIMOS observations, not included in the final sample. The small filled pink circles show the subset of $z\sim 1.5$ galaxies with two lines detected at $\SN>3$ and the final sample, the ``high-z sample'' is shown by the filled red circles. The mean errors on the $z\sim 1.5$ data are indicated by the black error bars at the top left. Regions of \Mstar\ and SFR containing 67, 95, and 99\% of the low-z star-forming catalogue from which the matched samples are drawn are indicated by the blue contours. The solid blue, dashed blue and dashed red lines represent the ``best, mixed'' main sequence fits at $z=0, \, 0.3$ and $ 1.5$, respectively, derived by Speagle et al. (2014). The pink filled region indicates the uncertainty the main sequence fit at $z~1.5$, based on the errors on the fitting coefficients derived by Speagle et al. (2014). Note that the high-z data represented by the three filled red circles just below the Speagle fit have been included in the final sample, as the SFR and \Mstar\ still consistent with the main sequence within the uncertainties.}
            \label{fig:ms}
        \end{figure}

        To increase the signal-to-noise ($\SN$) of the strong emission line fluxes and allow for robust metallicity and ionisation parameter determinations we rely on a stacked analysis (see Section \ref{sec:stacking}). Constructing stacked spectra requires reliable spectroscopic redshift estimates. We therefore require at least two emission lines, detected at a $\SN>3$, to confirm the spectroscopic redshift (to within $\Delta z < 0.005$). In all cases we require the presence of \Ha\ to confirm the redshift estimate. We remove 11 galaxies that do not meet these $\SN$ constraints from our sample.  We also exclude three galaxies which lie well below the star-forming main sequence at $z\sim 1.5$ (shaded pink region of Fig. \ref{fig:ms}) because our stacked analysis is aiming to probe properties that are representative of the dominant population of star-forming galaxies at $z\sim 1.5$. To reliably measure the Balmer decrement, \Ha/\Hb, and metallicity based on stacked data, we remove a further 18 galaxies for which the \Hb\ or $\OIII\lambda5007$ lines are completely masked by a skyline.  We refer to the final sample of 50 galaxies as the ``high-z sample''. 

        We rely on a combination of photometry and spectroscopy to derive the global properties referred to throughout this work. The stellar masses of the high-z sample are taken from the latest COSMOS photometric catalogue \citep{2016ApJS..224...24L}. These stellar masses are derived by fitting model spectra to the spectral energy distributions via \texttt{LePHARE} \citep{2011ascl.soft08009A} following the methods outlined by \cite{2015A&amp;A...579A...2I}. To convert the stellar masses to a Kroupa IMF we apply a constant scaling factor of $1.06$ \citep{2012ApJ...757...54Z}. The SFRs (and sSFRs) of our $z\sim 1.5$ sample were estimated from the dust-corrected \Ha\ luminosities using the conversions in \cite{2011ApJ...737...67M} and \cite{2011ApJ...741..124H} (consistent with a Kroupa IMF).

        We compare the \Mstar\ and \SFR\ of the high-z sample to both the full sample of galaxies observed with FMOS and DEIMOS and the sample for which two lines are detected at $\SN>3$ in Fig. \ref{fig:ms}. The final high-z sample (filled red circles) is consistent with the larger parent sample (open and small pink circles) but restricted to a narrower \Mstar\ and SFR range and limited to the star-forming main sequence at $z\sim 1.5$ by selection. As in previous high redshift studies \citep[e.g.][]{2016ApJ...816...23S,2014ApJ...785..153M,0004-637X-801-2-88}, our sample exhibits significantly higher SFRs and sSFRs than the dominant population of local ($z<0.1$) star-forming galaxies (solid blue line) although there is significant scatter in the \Mstar\ vs SFR relation.
    

    \subsection{Low-z comparison samples} 
        \label{sub:low_redshift_comparison_samples}

        We derive our low redshift (low-z) comparison samples from the Sloan Digital Sky Survey \citep[SDSS,][]{2000AJ....120.1579Y} Data Release 7 \citep[DR7,][]{2009ApJS..182..543A} catalogue.  We use the emission-line measurements and SFRs from the MPA-JHU catalogues \citep{2003MNRAS.346.1055K,2004MNRAS.351.1151B,2004ApJ...613..898T}. Like the high-z sample, SDSS SFRs are estimated from the \Ha\ luminosities after correcting for aperture loss of the SDSS fibers and dust extinction based on $\Ha/\Hb$, and are based on a Kroupa IMF. In order to perform a fair comparison with the high-z sample we adopt the stellar masses derived by \cite{2013ApJ...771L..19Z} using \texttt{LePhare}. Stellar masses derived using \texttt{LePhare} are $\sim 0.2\dex$ lower on average than those from the MPA-JHU catalogue, with a $\sim 0.2 \dex$ dispersion \cite{2013ApJ...771L..19Z}. As for our FMOS sample we normalise the SDSS stellar masses to a Kroupa IMF, such that the IMFs used for the SFR and \Mstar\ are consistent. We note that our choice of LePhare stellar masses yields results in a mass-metallicity (MZ) relation consistent with \cite{2013ApJ...771L..19Z} and that different sources of \Mstar\ and sSFR may impact the slope and turnover of the derived MZ relation.

        \begin{figure*}
              \begin{center}  
                    \includegraphics[width=0.85\textwidth,trim={1.0cm 1.5cm 1.5cm 0.5cm},clip]{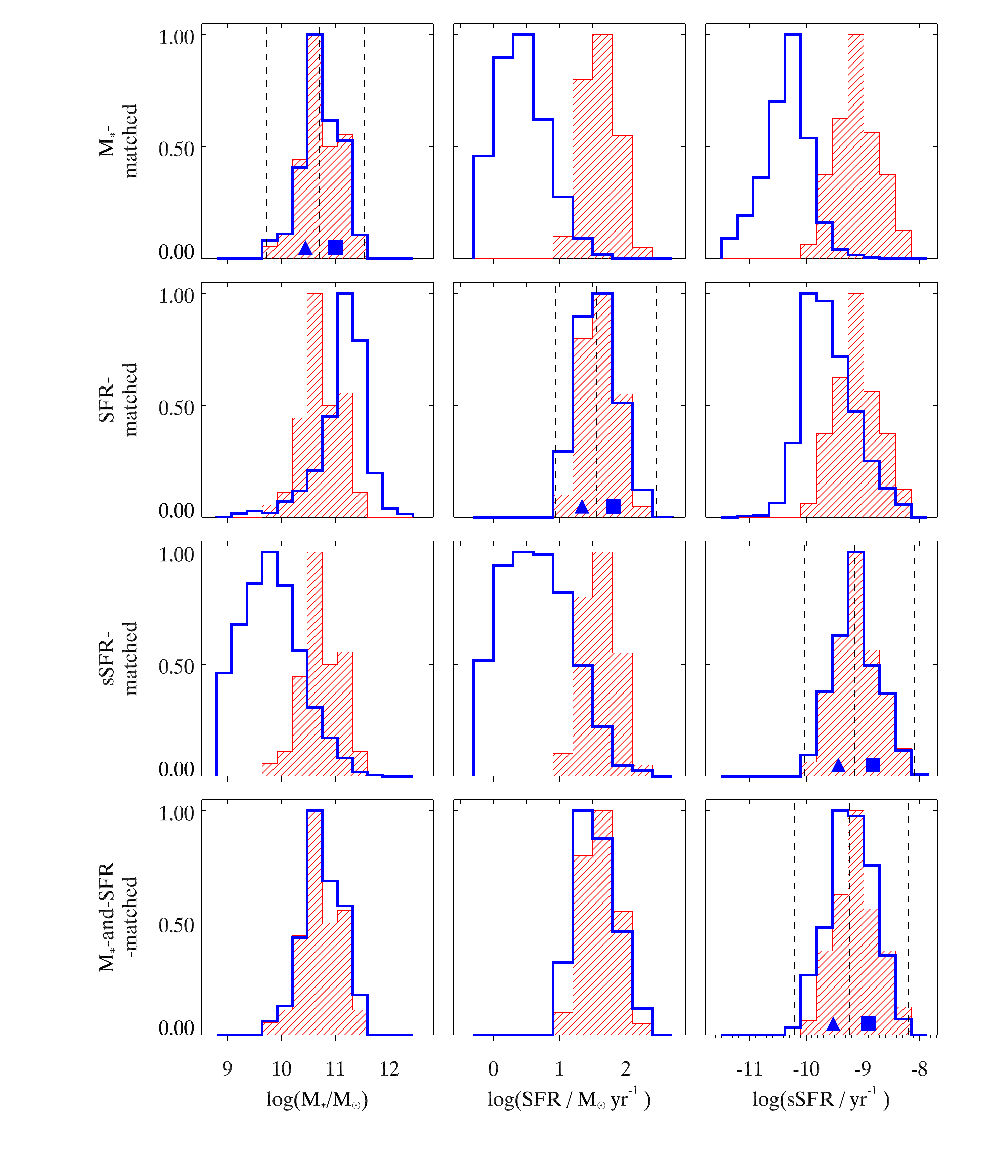} 
                  \end{center}
                \caption{Distribution of global properties for the high-z (red line fill) vs low-z matched (blue outline) samples.  The left, middle and right hand column show the distribution of \Mstar, SFR and sSFR respectively. From top to bottom, the blue outlined histograms in each row correspond to the \Mstar-, SFR-. sSFR- and \Mstar-and-SFR- matched low-z samples (as labelled on the left axis). The upper and lower bounds of the bins used in the stacked analysis are indicated by the dashed vertical lines whereas the mean value of the two low-z bins for each matched sample are indicated by the filled triangle and square respectively. Note that the triangle represents the half with the smallest values of the global property of interest whereas the square (with the larger number of sides) represents the bin with the larger values.}
                \label{fig:match_global}
        \end{figure*}

        To construct the catalogue from which we select our low-z comparison samples we apply further constraints. We discard galaxies for which the \Mstar\ or SFR are not constrained (i.e. negative or infinite value). To individually determine the metallicity and ionisation parameter and investigate the extent of AGN contamination, we enforce a $\SN$ constraint of $\SN>3$ on $\OII\lambda3727, \Hb, \Ha$ and $\NII\lambda6584$. However, we apply no $\SN$ constraint on $\OIII\lambda 5007$, to avoid biasing the average metallicities of galaxies with high stellar mass \citep{2012A&amp;A...547A..79F} and allow for a comparison with previous studies \citep[e.g.][]{2013ApJ...771L..19Z}. To reduce systematic errors from aperture effects we select galaxies at $z>0.04$ \citep{2005PASP..117..227K}. 

        We reject AGN based on the standard optical line ratios as for the high-z sample, by applying the \cite{2001ApJ...556..121K} maximum starburst criteria. The \cite{2001ApJ...556..121K} criteria places a theoretical upper limit on the location of star-forming galaxies beyond which the \OIII/\Hb\ and \NII/\Ha\ ratios can no longer be modelled without the contribution of an AGN. As discussed by \cite{2003MNRAS.346.1055K}, the application of the \cite{2001ApJ...556..121K} criteria to local galaxies may underestimate the contribution of AGN. We ensure that our choice of AGN cut-off had no impact upon our conclusions, by repeating our analysis using the upper limit of \cite{2003MNRAS.346.1055K}, which represents the maximum position of local star-forming galaxies on the BPT diagram. 

        We select four ``matched'' SDSS comparison samples from the ``low-z star-forming catalogue'', in order to study the evolution of ISM properties. These samples are matched to our high-z sample according to a different combination of global properties as follows,
        \begin{enumerate}[leftmargin=*]
             \item \Mstar-matched: matched only in \Mstar, 
             \item SFR-matched: matched only in SFR, 
             \item sSFR-matched: matched only in sSFR, and,
             \item \Mstar-and-SFR-matched: matched in both \Mstar\ and \SFR.
         \end{enumerate}  
        To capture a sufficient sample of low-z galaxies with SFRs (and \Mstar) equivalent to those of the high-z sample ($>10$ each) we employ an upper redshift cut-off of $z<0.3$. We refer to the sample of 224 892, star-forming low-z ($0.04<z<0.3$) SDSS galaxies from which we draw our comparison samples as the ``low-z star-forming catalogue''. The distribution of \Mstar\ and SFR for the low-z star-forming catalogue are indicated by blue contours in Fig. \ref{fig:ms}, which contain 67, 95, and 99\% of the sample. We recover an evolution in the main sequence from the low-z star-forming catalogue to the high-z sample (red filled circles) and show that the majority of $z<0.3$ star-forming galaxies encompass the same stellar mass range as the high-z sample, but typically have lower SFRs.

        Each ``matched'' low-z sample is created by randomly selecting an ensemble of $z<0.3$ counterparts for every high-z galaxy. To ensure that the distribution of the global property being matched is the same for the low- and high-z samples, we select the same number of low-z counterparts for each high-z galaxy. We limit the \Mstar- and sSFR-matched samples to 100 low-z counterparts per high-z galaxy to minimise the computational cost. Further increases in the size of these matched ensembles had no impact upon the mean or spread in the emission-line properties of the stacked samples. For the two samples matched in SFR, the size of the matched ensembles is determined by the minimum possible number of matches to galaxies in the high-z sample. Because we find only 37 low-z galaxies, with equivalent SFRs to the highest SFR galaxy in the high-z sample, we select 37 counterparts for each of the galaxies in the the high-z sample. Similarly, we select only 10 counterparts in \Mstar\ and SFR, per high-z galaxy, to construct the \Mstar-and-SFR-matched sample.

        The tolerances used for matching are selected based on the mean errors on the high-z sample. We create the \Mstar-matched sample by selecting galaxies for which the \Mstar\ is consistent to within 0.1\dex\ with the \Mstar\ of the high-z counterpart and create the SFR-matched sample by selecting low-z galaxies for which the SFR is consistent to within $0.2\dex$ that of each high-z galaxy. To create the sSFR-matched sample, we select galaxies for which the sSFR is consistent to within 0.15\dex\ that of the high-z galaxy. Only the ratio of SFR to \Mstar\ is equivalent to the high-z counterparts, not the values of \Mstar\ or SFR (see third row of Fig. \ref{fig:match_global}). For our final sample we select local counterparts for which both the \Mstar and SFR are consistent to within $\sim 0.2\dex$. 

        We compare the distributions of \Mstar, SFR and sSFR of our low-z matched samples to the high-z sample in Fig. \ref{fig:match_global}. Because of the evolution of the main-sequence and the sample characteristics of SDSS, the majority of \Mstar-matched, low-z galaxies have both lower SFRs and sSFRs than the high-z sample (top row of Fig. \ref{fig:match_global}). Conversely, the majority of SFR-matched low-z galaxies have higher \Mstar\ and lower sSFRs. Most sSFR-matched local galaxies have both lower \Mstar\ and SFRs than the high-z sample. 
    


\section{High-z data} 
    \label{sec:highz_data}

    \subsection{Observations and data reduction}
        \label{sub:data_reduction}

        Our sample of $z\sim 1.5$ galaxies is based on three sets of observations, two from FMOS (H-band and J-band) and one with DEIMOS. The H-long and H-short observations were conducted over fourteen nights in March 2012, January 2013, December 2013, January 2014 and February 2014 with follow-up J-band observations conducted over 6 nights in March 2012, January 2013 and February 2014. The average seeing for the FMOS observations was $\sim 1.0"$, although it varied significantly over these six nights. Approximately 940 galaxies at $z\sim 1.5$ were observed in the H-band, with a subset of $\sim 380$ observed in the J-band. 

        The corresponding DEIMOS observations were conducted over two clear nights, UTC February 24 and 25, 2014. All DEIMOS observations were conducted with the 600ZD grating centred at 7500\ang, the OG550 filter and 1" slit width. The average seeing over the two nights of the DEIMOS observations was $\sim 0.75"$.  Approximately 430 galaxies at $1.4<z<1.7$ were observed with DEIMOS over the course of the two nights. The FMOS and DEIMOS observations are described in more detail in \cite{2015ApJS..220...12S} and \cite{2017MNRAS.465.3220K} respectively.

        The raw FMOS-COSMOS and COSMOS-\OII\ science frames were reduced as described in \cite{2015ApJS..220...12S} and \cite{2017MNRAS.465.3220K} respectively. Both surveys made use of publicly available reduction pipelines to perform bias removal, flat fielding, cosmic ray rejection and slit-tilt corrections. Wavelength calibrations were performed on a slit-by-slit basis, based on standard arc lamp images. FMOS spectra were obtained using the publicly available FMOS Image-Based Reduction Package (\texttt{Fibre-pac}; Iwamuro et al. 2012). The DEIMOS science frames were processed via the publicly available IDL based pipeline, \texttt{spec2d}, developed by the DEEP2 survey team \citep{2012ascl.soft03003C,2013ApJS..208....5N} and reduced to 1D by calculating the total flux over the effective aperture, as described in \cite{2017MNRAS.465.3220K}. The initial flux calibration of each (wavelength calibrated) spectrum was performed using flux standard stars observed along with the other scientific targets \citep[see][]{2015ApJS..220...12S,2017MNRAS.465.3220K}. These initial flux calibrations did not account for variable seeing conditions and atmospheric throughput.  
    

    \subsection{Individual spectral analysis} 
        \label{sub:ind_spectral_analysis}

        Our study is based mainly on a stacked analysis of the FMOS and DEIMOS spectra (see Section \ref{sec:stacking}). However, we rely on individual spectra to derive the SFRs of our sample and show the emission-line properties, metallicities and ionisation parameters for the subset of galaxies with the highest $\SN$ in the relevant emission lines in Figures \ref{fig:BPT} to \ref{fig:qz_MandSFR}. We separately derive the spectroscopic redshifts and emission-line fluxes of the two surveys. The fitting procedures for the FMOS and DEIMOS data are described in detail in \cite{2015ApJS..220...12S} and \cite{2017MNRAS.465.3220K} respectively. Most FMOS spectra lack stellar continuum detections and are dominated by the strongest emission lines. Because the majority of spectra lack significant detections of the stellar continuum, we used linear functions to approximate the median flux level of pixels adjacent to the emision line of interest and subtract this component prior to fitting emission lines. 

        We account for the effects of stellar Balmer absorption and dust extinction. To correct for \Hb\ absorption we apply a correction factor based on the relationship derived by \cite{0004-637X-792-1-75} for the FMOS data (discussed in Section \ref{sub:stacking_high_z_bins}). We do not correct the \Ha\ flux for stellar absorption because the flux loss is negligible ( $<2\%$ \citealt{2013ApJ...777L...8K}). We correct for the effects of dust extinction by inferring a reddening correction from the measured Balmer decrement, $\Ha/\Hb$ and applying the \cite{1989ApJ...345..245C} extinction curve with $R_v = 3.1$. For galaxies with reliable Balmer decrements ($\SN >3$ for both \Ha\ and \Hb) we estimate the level of extinction via the measured Balmer decrement and propagate the uncertainty on the Balmer decrement to the corrected line fluxes. In cases where the Balmer decrement is unreliable (either $\SN (\Hb)<3$ or $F_{\lambda_\Ha,0}/F_{\lambda_\Hb,0}<2.86$), we use upper (lower) limits on the corrected fluxes based on the upper (lower) limits of the Balmer decrements available for our sample. 


    \subsection{Relative aperture corrections} 
        \label{sub:aperture_corrections}

        To compare the relative fluxes from different wavelength bands, we must account for the variation in aperture size between the two surveys as well as the variation in seeing conditions on different nights. We therefore derive three separate aperture correction factors for each galaxy, corresponding to H-band (\Ha\ and \NII), J-band (\Hb\ and \OIII) and DEIMOS (\OII) observations. All aperture correction factors are derived using the Hubble Space Telescope (HST)/Advanced Camera for Surveys (ACS) I$_{\mathrm{F814W}}$-band images, smoothed according to an effective seeing. We thereby assume that the rest-frame UV and line emission have a similar spatial distribution. This assumption is partly justified by the tight correlation between the half-light radii in $\Ha$ and the ACS I-band for the SINS/zCSINF galaxies \citep[e.g.][]{2011ApJ...743...86M}. 
        The derivation of aperture correction factors for the FMOS H- and J-band observations is described in detail in \cite{2015ApJS..220...12S}. We apply a similar method to derive the \OII\ aperture correction factors, which we define as the ratio between the total UV flux of each galaxy and the portion of UV flux sampled by the DEIMOS slit. For each galaxy we define a rectangular region encompassing the entire galaxy on the HST/ACS mosaic and take the total UV flux to be the sum of the UV flux over all pixels in this manually defined region.  To take into account the increased size of the galaxy due to the seeing conditions, we convolve each HST/ACS I-band image with a Gaussian point spread function (PSF) matching the seeing conditions at the time the DEIMOS observation was conducted. We determine the amount of UV flux sampled by the DEIMOS slit by summing the UV flux over the portion of the galaxy covered by the slit. The area covered by the slit is defined by the DEIMOS slit dimensions ($1"\times 5"$), the central RA and DEC, the position angle of the slit and the bounds of the galaxy itself. Based on the impact of seeing variations, the effect of the PSF used and positioning errors we apply a uniform uncertainty of $\pm 0.05$ to all aperture correction factors.

        The three aperture correction factors (\OII, J-band and H-band) vary significantly. The mean \OII, J-band and H-band corrections factors are 1.4, 3.7 and 2.3 respectively, for the $z\sim 1.5$ sample for which we have FMOS H- and J-band and DEIMOS observations. The difference between the aperture correction factors derived for each survey is the result of two factors; the larger physical size of the DEIMOS slit compared to the FMOS fibre and better seeing conditions for the DEIMOS observations (average of $0.75"$ vs $0.9"$). The effects of poorer seeing conditions are also apparent when comparing the two sets of aperture correction factors derived for the FMOS data. Because J-long observations were often conducted under poorer seeing conditions, most J-band correction factors for our sample are greater than the corresponding H-band correction factors. 

       	We scale both the emission-line fluxes of individual galaxies and the spectra used for stacking by the derived aperture corrections. The relative size of the three aperture corrections means that their application may have a significant impact on the extinction corrections and individual line ratios. We ensure that the relative aperture correction factors do not bias individual emission-line flux ratios. However, we note that there is an inherent dispersion in the corrections that needs to be considered when addressing issues such as the difference in dust extinction and the width of the star-forming sequence (Fig. \ref{fig:ms}). 


\section{Stacking}
	\label{sec:stacking}

    We characterise the ISM properties of the matched low- and high-z samples based on the emission-line fluxes of stacked spectra. Of the 13 high-z galaxies for which all five lines (used to diagnose the metallicity and ionisation parameter) are detected at $\SN>3$, only six have sufficient $\SN$ in the required lines after applying aperture and extinction corrections to confidently infer the ionisation parameter and metallicity (see Section \ref{sec:diagnosing}). The ISM properties of $z\sim 1.5$ star-forming galaxies are not sufficiently represented by this subset of six galaxies, which exhibit a limited range of \Mstar\ and SFR, a significant variation of inferred ISM properties and an inherent bias towards the properties of galaxies with more luminous emission lines. 

    By using stacked spectra, we increase the strength and $\SN$ of the strong emission line fluxes of the sample, reduce the scatter introduced by aperture and dust corrections and include galaxies with non-detections in some of the emission lines. The higher $\SN$ in all five emission lines of the stacked spectra allows us to diagnose the average ionisation parameter and metallicity with greater precision. By performing a stacked analysis we assume that the single artificial galaxy spectrum created by stacking is representative of the average properties of the sample being stacked. We investigate the validity of this assumption in Section \ref{sub:validity}. 

    \begin{table}
            \centering
            \ra{1.3}
            \caption{Binned and stacked samples}
            \label{tab:bins}
             \begin{tabular}{L{1.5cm}C{1.2cm}C{1.2cm}C{1.2cm}C{1.2cm}}
             \toprule
              Sample  & Sample size & Matched property  & Binned property   & Number per bin  \\
             \toprule
                High-z & 50 & - & \Mstar & 25  \\
                High-z & 50 & - & SFR    & 25  \\
                High-z & 50 & - & sSFR   & 25  \\
            \toprule
                Low-z,  \Mstar-matched           & 5000  &  \Mstar          & \Mstar  & 2500  \\
                Low-z,  SFR-matched              & 1850  &  SFR             & SFR     & 925   \\
                Low-z,  sSFR-matched             & 5000  &  sSFR            & sSFR    & 2500  \\
                Low-z,  \Mstar-and-SFR-matched   & 500   &  \Mstar\ and SFR   & sSFR    & 250   \\
             \bottomrule
           \end{tabular}
    \end{table}

    \subsection{Binning} 
        \label{sub:binning_samples}

        To create the stacked spectra we categorise the galaxies in each sample according to the global property of interest (\Mstar, SFR or sSFR). For each global property, we separate the galaxies in our high-z sample into two equally sized bins, representing the halves with the low and high values of this global property. We use two bins for each high-z sample because a greater number of bins led to a significant scatter in the emission-line properties of different bins due to larger errors associated with smaller sample sizes. The final results are stable with slight changes in binning when we consider two bins for each sample.

        Each matched, low-z sample is also equally separated into two bins according to the property in which it was matched to the high-z sample. The samples and bins are given in Table \ref{tab:bins}. Note that the \Mstar-and-SFR-matched low-z sample is separated into two bins of sSFR to allow for a two-dimensional visual comparison with the high-z sample. We indicate the mean and the range of the matched global property spanned by each bin in Fig. \ref{fig:match_global}. 


    \subsection{Stacking high-z bins} 
        \label{sub:stacking_high_z_bins}

        We create three stacked spectra for each bin of the $z\sim 1.5$ galaxies, one each for the H-band, J-band and DEIMOS spectra. Our method is similar to that of previous FMOS studies that rely on a stacked analysis \citep[see][]{0004-637X-792-1-75,2017ApJ...835...88K}. As noted in both these FMOS studies, a significant fraction of pixels in each FMOS spectrum are impacted by strong residual sky lines. These pixels are indentified from the noise spectra and masked out before stacking.  Individual spectra are then converted to the rest-frame, based on their spectroscopic redshifts, and are resampled onto a common wavelength grid. We choose wavelength spacings corresponding to the observed frame spectral resolution of FMOS (1.25 \ang/pixel) and DEIMOS (0.65 \ang/pixel), for galaxies at $z\sim 1.5$. Thus, the wavelength spacings for the FMOS and DEIMOS spectra are 0.5\ang/pixel and 0.26 \ang/pixel, respectively.

        We do not fit the underlying stellar continuum or absorption, because of the lack of continuum detections in the high-z spectra. Instead, we subtract a single value for the stellar continuum for each spectrum based on the median flux in each band. The de-redshifted, resampled and continuum-subtracted spectra are scaled by their corresponding aperture correction factors (see Section \ref{sub:aperture_corrections}). To create the final, stacked spectrum for each bin we determine the mean spectrum using \texttt{IDL}'s \texttt{resistant\_mean.pro}, available from the Astronomy User's Library. Values that deviate from the median value at each pixel by more than five times the absolute deviation are removed via a $5\sigma$ clipping scheme.

        The relative calibrations used while stacking are likely to be a large source of error. However, the application of aperture corrections factors after continuum subtraction mean that we also rely heavily on the accuracy of the flux calibration of each band. We note that the absolute flux calibration, especially of the J-band spectra, may be a large source of error in the stacked spectra, leading to the selection of fewer bins of spectra to stack. 

        \begin{figure}
          \centering
                \includegraphics[width=0.5\textwidth,trim={1.0cm 0.8cm 1.0cm 1.5cm},clip]{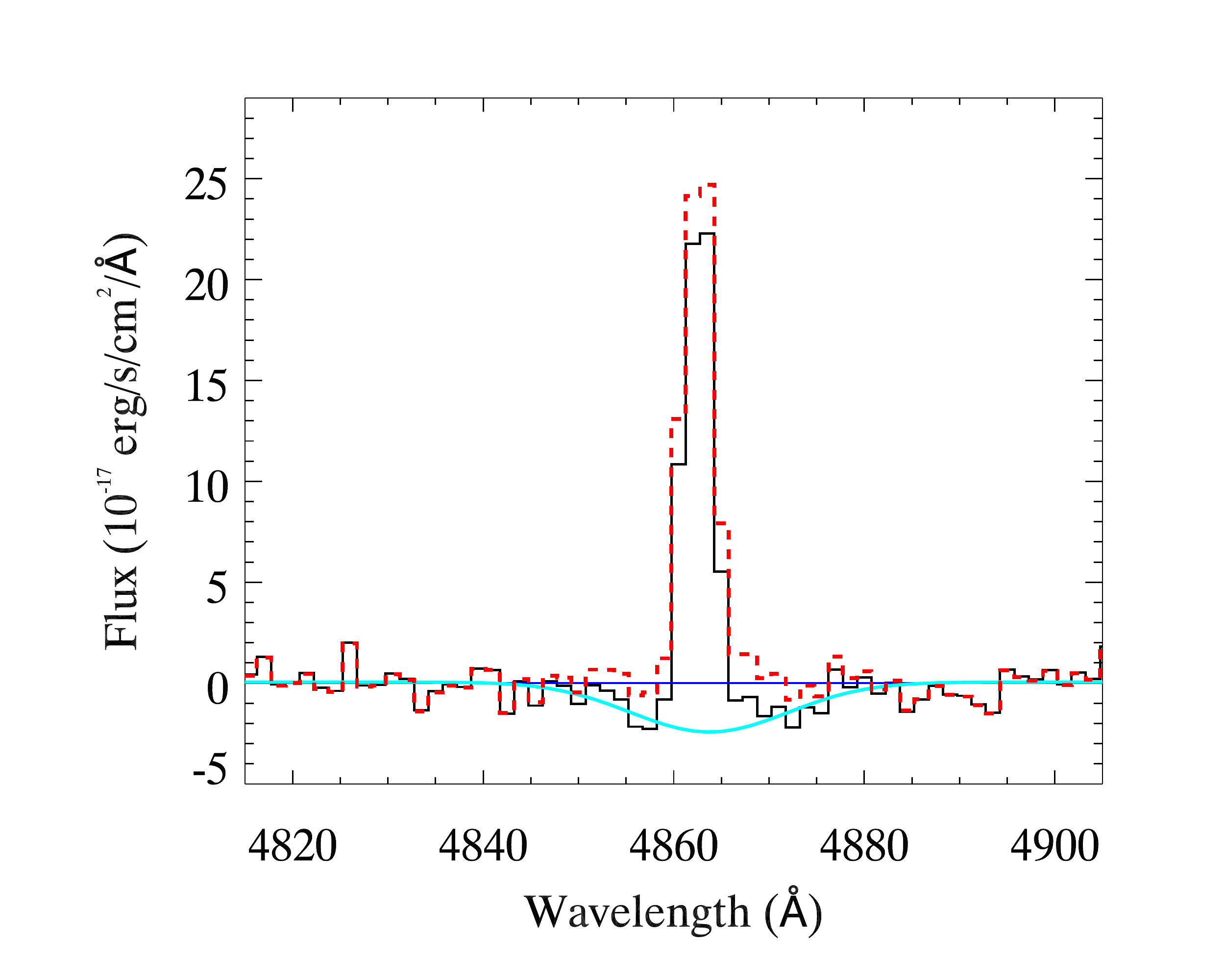} 
            \caption{Example of the effect of Balmer absorption on a single low-z galaxy spectrum. The fit to the Balmer absorption (solid cyan line) is added to the measured spectrum (solid black line) resulting in a spectrum with a larger \Hb\ emission feature (red dashed line). 
            }
            \label{fig:balmerabs}
        \end{figure} 

        \begin{figure*}
              \begin{center}  
                    \includegraphics[width=0.7\textwidth,trim={0cm 0.5cm 0.5cm 0.5cm},clip]{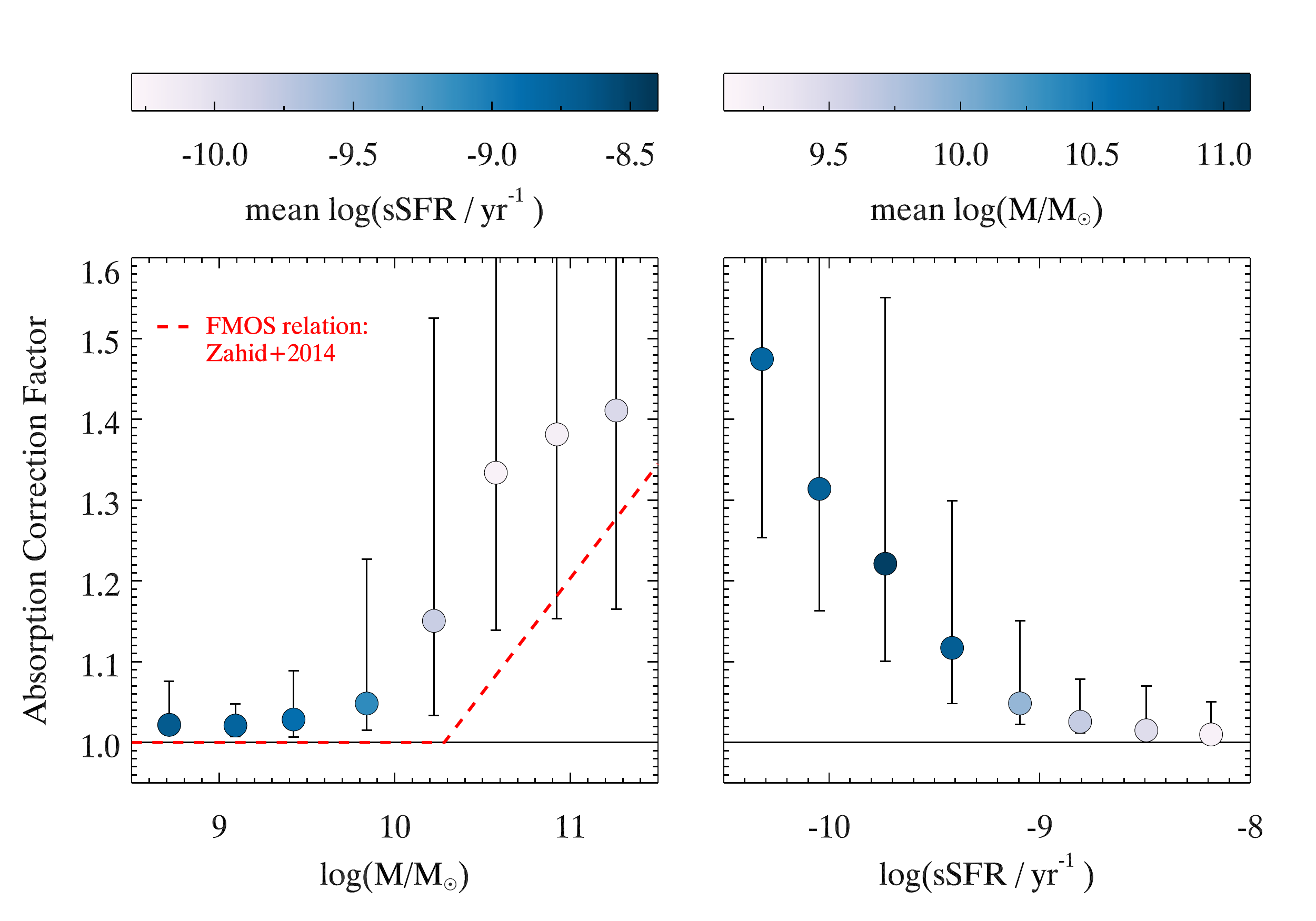} 
              \end{center}
            \caption{Balmer absorption correction factor ($F(\Hb_\mathrm{corr})/F(\Hb)$) per bin of \Mstar\ (left panel) and sSFR (right panel) for galaxies in the four low-z matched samples. Filled circles indicate the mean absorption correction factor per bin, with errors bars extending to the 16th and 84th percentile values. The mean sSFR of the \Mstar\ bins in the left hand panel is indicated by the shading described in the top left colourbar whereas the mean \Mstar\ of the sSFR bins in the right hand panel is indicated by the shading described in the top right colour bar. The corrections factors derived for the low-z matched samples are compared to the relationship derived for the $z\sim 1.5$ FMOS sample of Zahid et al. (2014), (equation \eqref{eq:bac}, dashed red line). }
            \label{fig:balmer_corr_comp}
        \end{figure*} 
        
        We do not correct the individual high-z spectra for Balmer absorption because of the lack of \Hb\ detections. However, \cite{0004-637X-792-1-75} show that the Balmer absorption of FMOS spectra becomes non-negligible for high \Mstar\ ($\Mstar > 10^{10.5} \Msun$) galaxies.  We therefore scale the emission-line fluxes of the stacked spectra by the absorption correction factors, $f_{\mathrm{corr}}$, predicted for the FMOS sample,
         \begin{align}
            f_{\mathrm{corr}} =  \mathrm{max} \left\{ 1, 1.09 + 0.30[\log(\Mstar/\Msun) - 10] \right \} \, .
            \label{eq:bac}
        \end{align}
        based on \Mstar\ of the stacked sample. We apply the mean correction factor for the stellar masses in each bin and account for the variation in correction factors by including the standard deviation of the correction factors when deriving the uncertainty of the \Hb\ flux. We note that our conclusions are unaffected by the Balmer absorption corrections applied to the high-z sample. 
    


        \begin{figure*}
          \centering
            \begin{minipage}{\linewidth} 
                \includegraphics[width=0.32\textwidth,trim={0.5cm 0.5cm 2.0cm 2.0cm},clip]{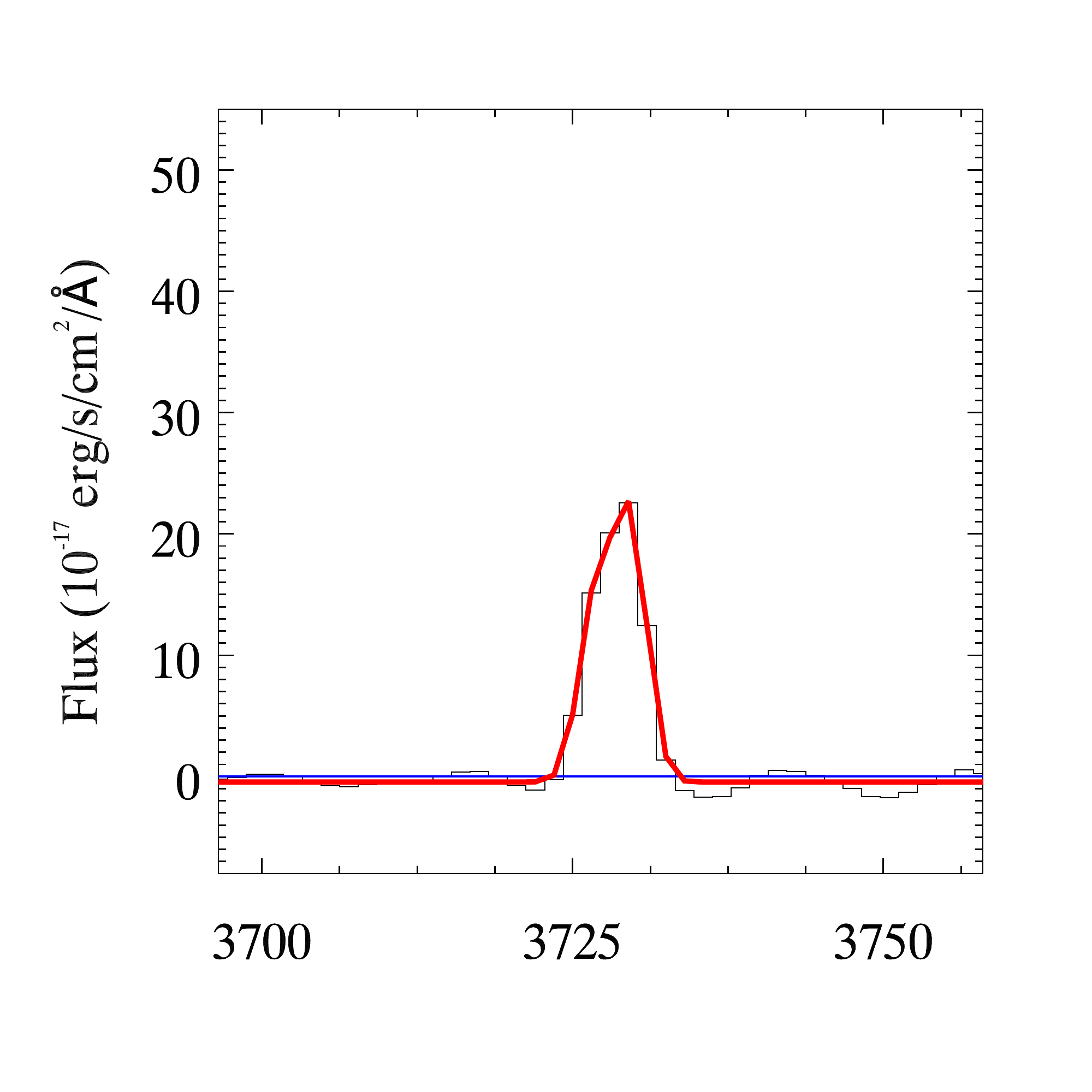} 
                \includegraphics[width=0.345\textwidth,trim={4.0cm 0.5cm 2.2cm 2.0cm},clip]{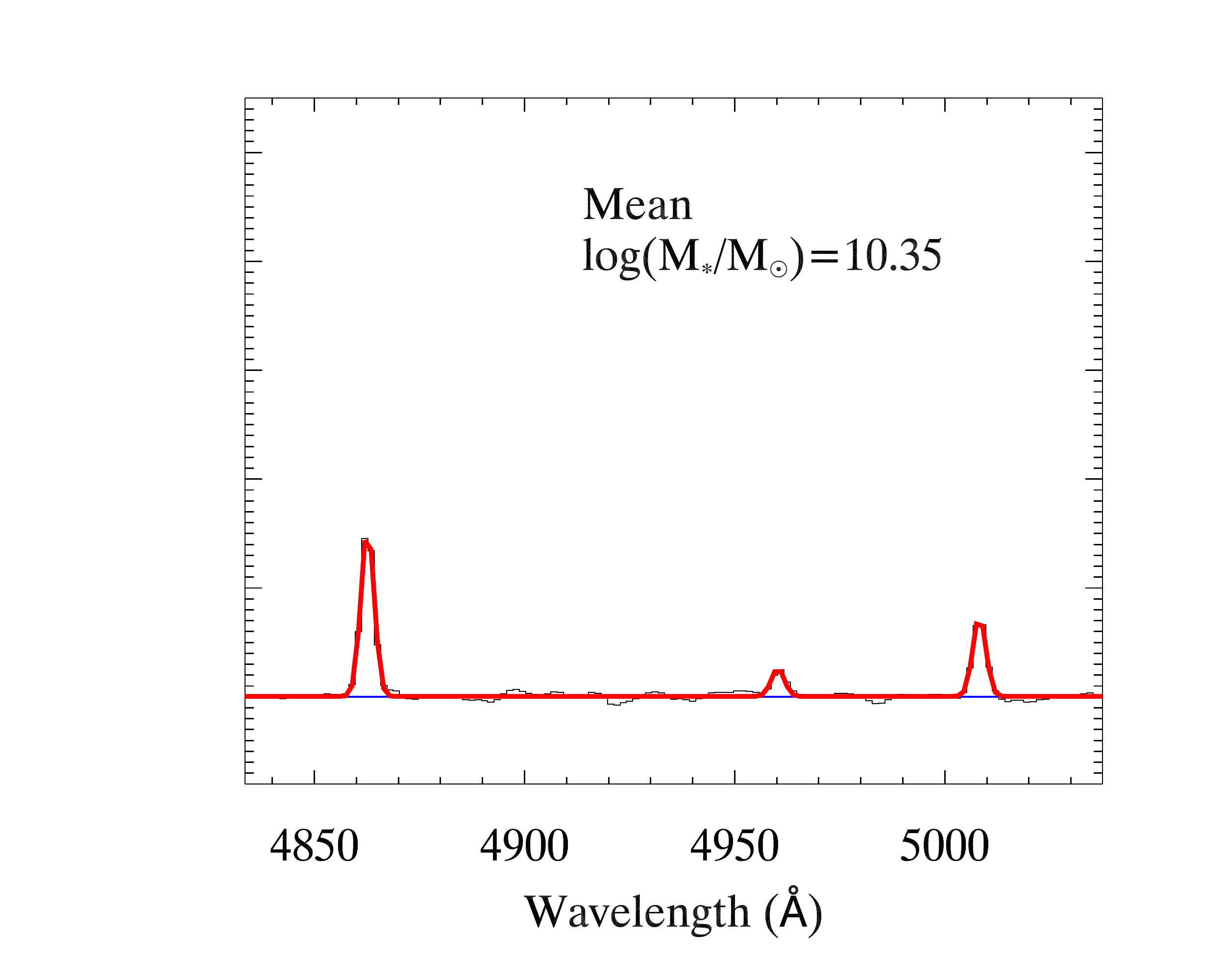}
                \includegraphics[width=0.305\textwidth,trim={4.0cm 0.5cm 2.0cm 2.0cm},clip]{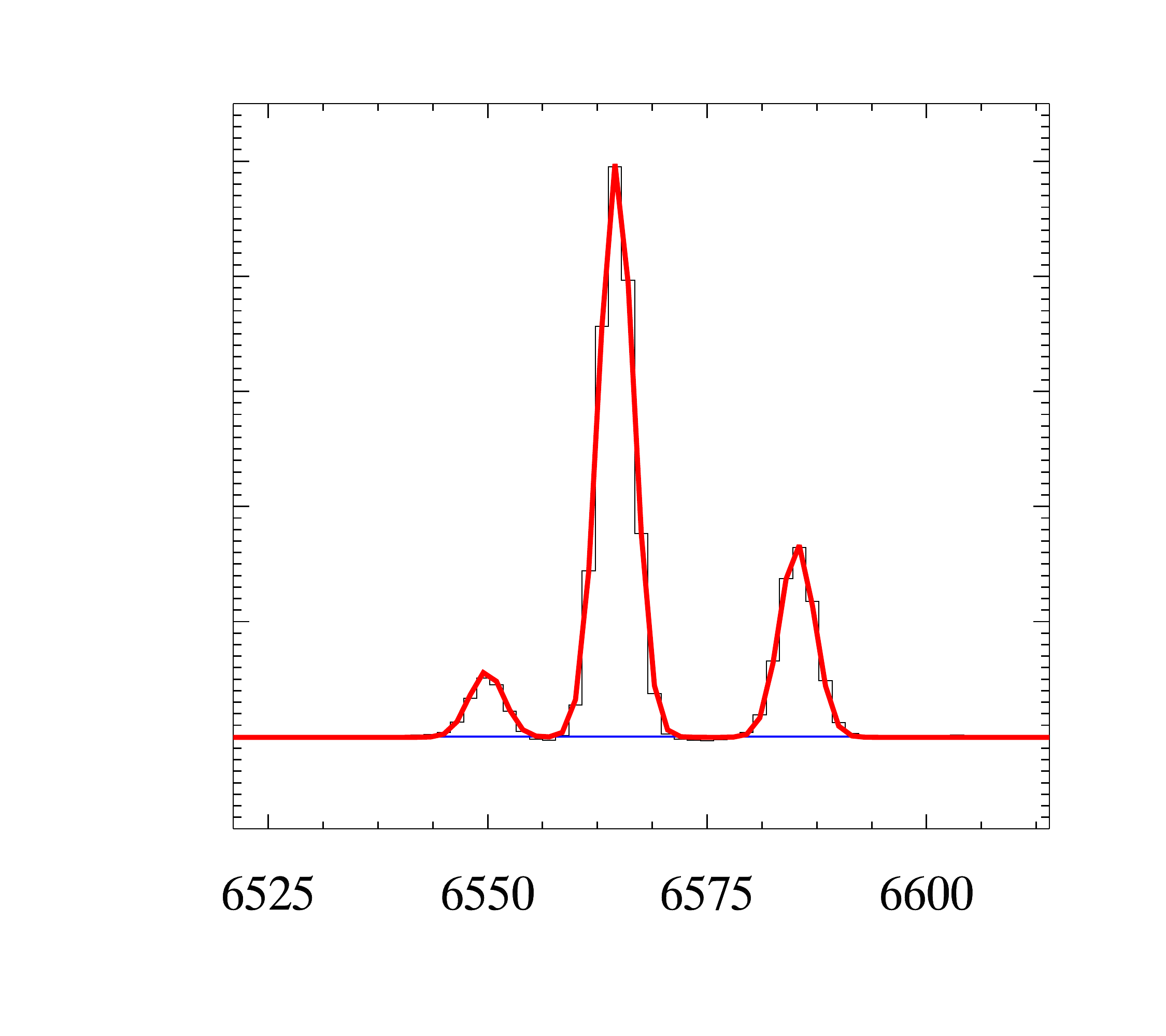}
                \\
                \includegraphics[width=0.32\textwidth,trim={0.5cm 0.5cm 2.0cm 2.0cm},clip]{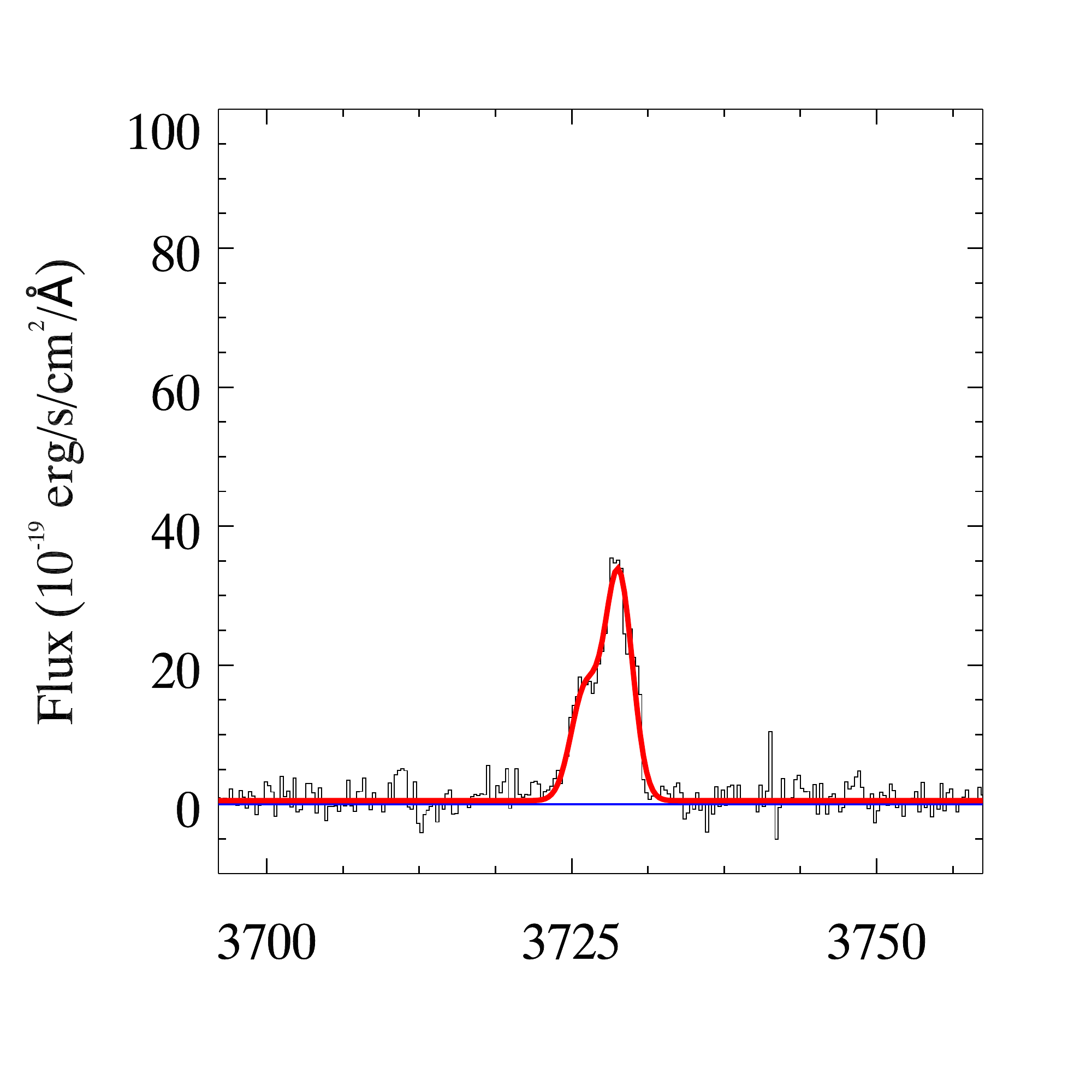} 
                \includegraphics[width=0.345\textwidth,trim={4.0cm 0.5cm 2.2cm 2.0cm},clip]{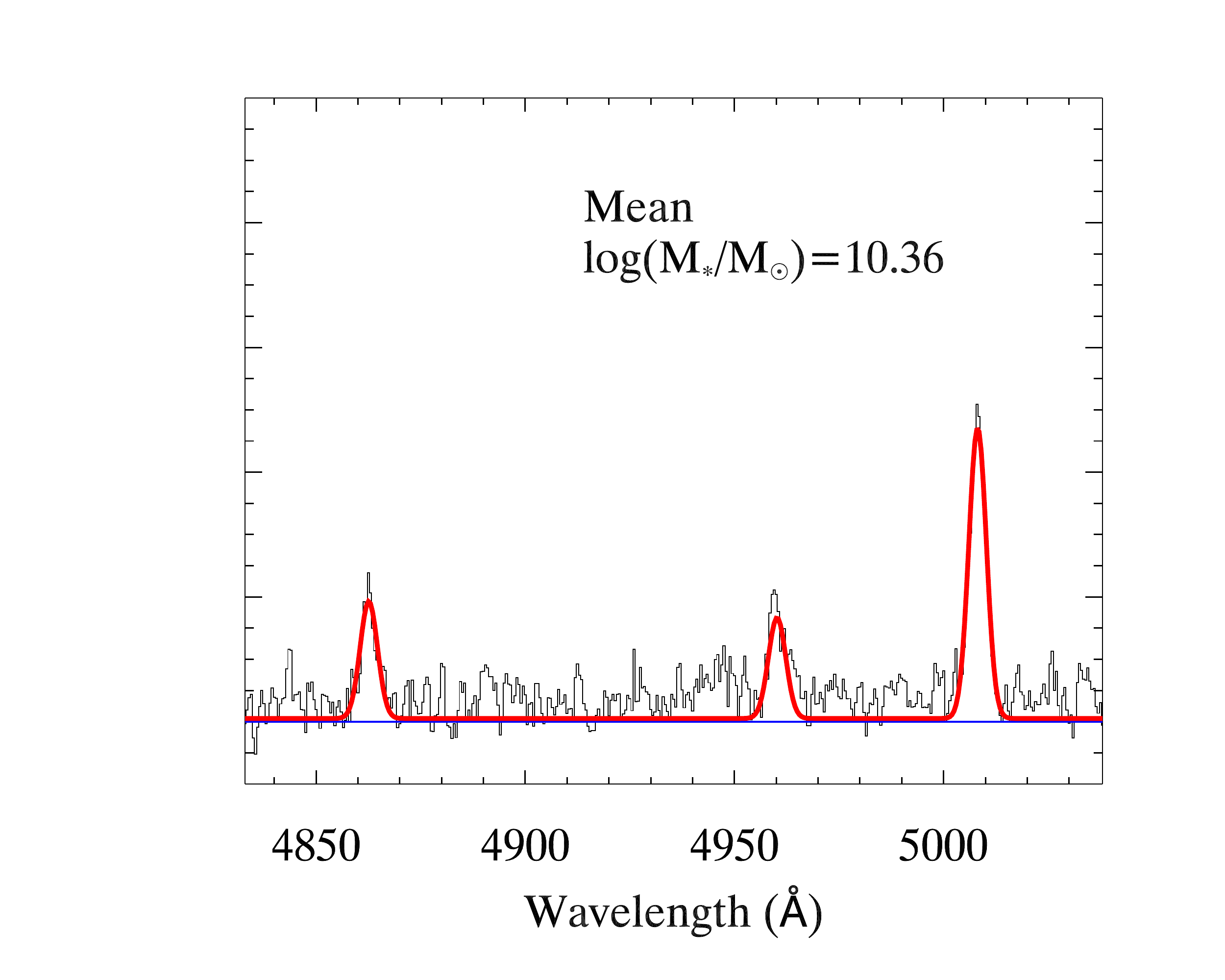} 
                \includegraphics[width=0.305\textwidth,trim={4.0cm 0.5cm 2.0cm 2.0cm},clip]{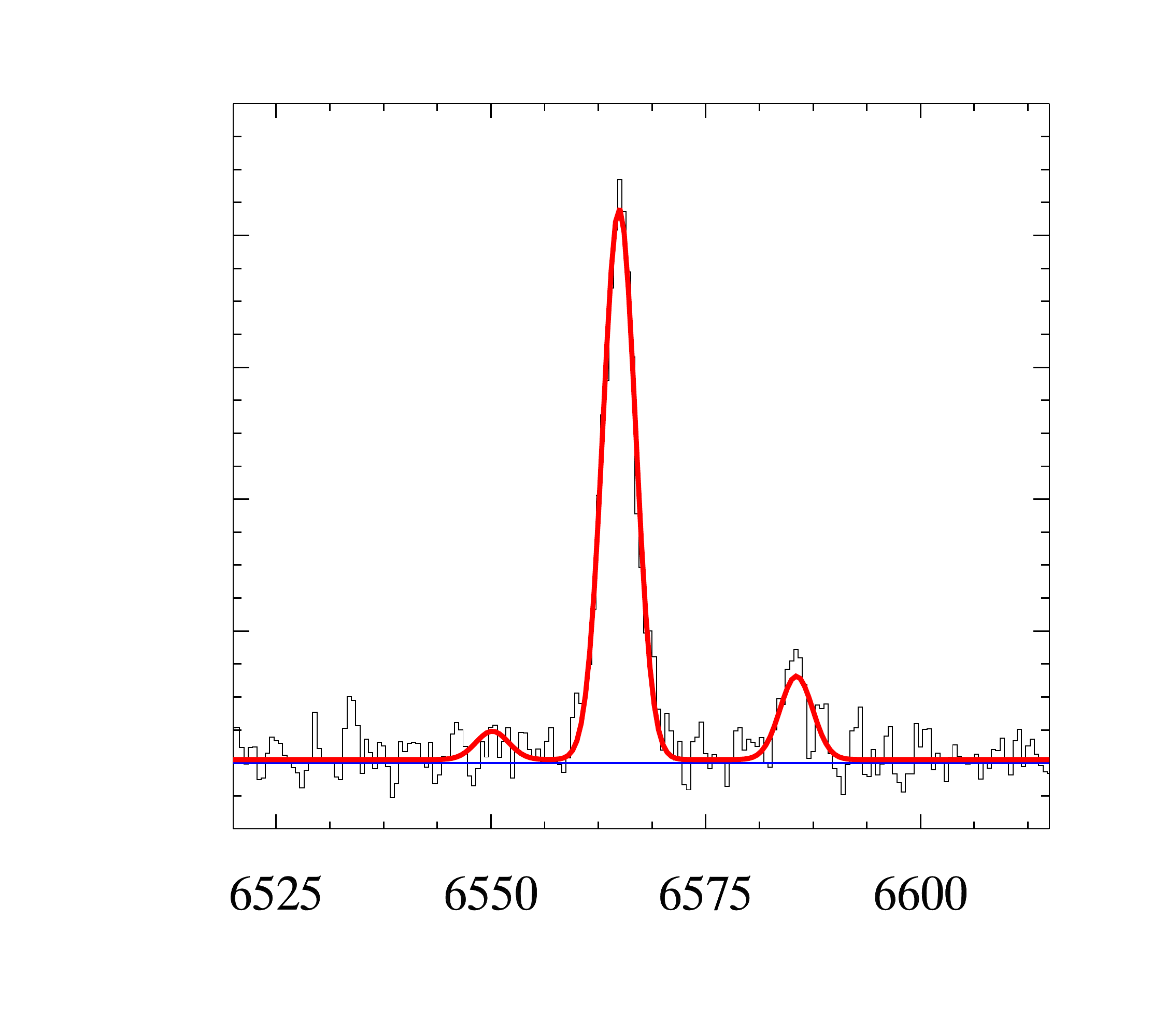}
            \end{minipage}
            \caption{Stacked spectra (solid black) of the lowest mass bin of the low-z \Mstar-matched (top row) and high-z (bottom row) samples compared to the emission line fits (solid red). The three panels in each row indicate the three fitting windows used to fit the five emission lines. Left panel: $\OII\llambda 3726,3729$ fit. Middle panel: \Hb\ and $\OIII\llambda4959,5007$. Right panel: $\NII\llambda 6548,6584$ and \Ha. }
            \label{fig:comp_spec}
        \end{figure*}


    \subsection{Stacking low-z bins} 
        \label{sub:creating_stacked_spectra}

        To create the stacked low-z spectra, we use the reduced and continuum subtracted 1D SDSS spectra from the SDSS DR7 catalogue \citep[see ][for details]{2002AJ....123..485S}. In contrast to the high-z sample, all five emission lines of the low-z spectra are captured within a single spectrum. Thus, we do need to account for any relative aperture covering fractions for different bands. We assume that our $z>0.04$ redshift cut-off ensures that the stacked spectra are not biased to only the central portions and create a single stacked spectrum per low-z bin. We investigate this assumption further in Section \ref{sub:validity}. 

        We do not mask out any pixels in the SDSS spectra because the emission lines are unaffected by residual skylines. Each SDSS spectrum is converted to the rest-frame wavelength based on the spectroscopic redshift.  Although the SDSS spectra have been continuum subtracted using a combination of stellar population synthesis models, most of the individual spectra in our sample contain significant residual Balmer absorption features. The stellar \Hb\ absorption of SDSS spectra is severely underestimated, especially for galaxies with low sSFR (see discussion in \citealt{2012MNRAS.419.1402G}). To remove the residual effects of the stellar absorption we fit a gaussian profile to the underlying absorption and add the profile to the spectrum as shown in Fig. \ref{fig:balmerabs}. Each spectrum is then resampled to a common wavelength grid of 1.5 \ang \citep[as in ][]{2014ApJS..213...35G}. Like the stacked high-z spectra, our stacked low-z spectra are created by taking the mean of the resampled and de-redshifted spectra for each bin via the ``resistant\_mean'' algorithm with a $5\sigma$ clipping scheme.      

        We show the extent of our corrections to the residual Balmer absorption for the low-z matched samples in Fig. \ref{fig:balmer_corr_comp}, via the absorption correction factor ($F(\Hb_\mathrm{corr})/F(\Hb)$).  The mean absorption correction factors of the low-z SDSS galaxies increase with \Mstar\ but scale inversely with sSFR. However, we note that there is a significant spread in the absorption correction factors of high mass ($\Mstar>10^{10.5}\Msun$) low-z galaxies. This spread is the result of the spread in sSFRs. High mass galaxies in only the sSFR-matched sample require significantly less correction for Balmer absorption than the majority of high mass low-z galaxies, which have lower sSFRs. In fact, the mean correction factors of the low-z, sSFR-matched sample are consistent with the relation derived for the FMOS sample (dashed red line), which exhibits the same range of sSFRs. The offset of the absorption correction factors of the entire matched low-z sample (filled circles) from the FMOS relation therefore reflects the lower sSFRs of most galaxies in the matched, low-z samples.


    \subsection{Emission-line fluxes of stacked bins} 
        \label{sub:emission_line_fluxes}

        We fit the emission lines of the low- and high-z stacked spectra in three separate wavelength regimes, around; (1) \OII,  (2) \Hb\ and $\OIII\lambda 5007$ and (3) \Ha\ and $\NII\lambda 6584$. 
        We use IDL's \texttt{mpcurvefit.pro} to fit the supplied gaussian functions. To extract the emission-line fluxes of the \OII\ doublet, we assume a double gaussian profile. For the \Hb\ and $\OIII\llambda 4959,5007$ fits, we apply a single Gaussian profile to each line and fix the relative amplitudes of the \OIII\ lines to the laboratory value of 2.887 \citep{Tachiev_FFischer_2001}. Similarly, we fit \Ha, $\NII\lambda6548$, $\NII\lambda6584$ with single Gaussian functions and fix the relative amplitude of the \NII\ lines to the laboratory value of 2.941 \citep{Tachiev_FFischer_2001,NIST_ASD}. We use a single FWHM to describe the widths of the lines in each wavelength regime but note that the values of FWHM may differ for the three wavelength regimes. We ensure that the emission-line fluxes of stacked spectra were consistent with the sum of the emission-line fluxes derived for individual FMOS spectra for subsamples consisting of spectra with all emission lines detected at $\SN>3$. The stacked spectra and corresponding emission-line fits of the lowest mass bin of low- and high-z stacked spectra are shown in Fig. \ref{fig:comp_spec}.  

        We estimate the errors on the emission-line fluxes of stacked spectra via a bootstrapping technique. For each stacked spectrum, we generate 1000 additional stacked spectra, each constructed from a random sample of galaxies drawn with replacement from the sample used for the actual stacked spectrum. We take the errors on the aperture corrections into account when constructing the additional spectra. To avoid biasing the results for the smaller bin sizes of the high-z sample, we ensure that no more than six of the 25 galaxies in each bin are replaced with duplicates. For each additional spectrum, we re-fit the emission lines as for the real spectrum. The errors on the measured emission-line fluxes are taken to be the standard deviation of the distribution of emission-line fluxes for the additional spectra. Our choice of the number of additional spectra and percentage to resample ensures that our errors are approximately gaussian.

        We correct the emission-line fluxes of the stacked spectra for extinction based on the Balmer decrement of each stacked bin of galaxy spectra.
        We assume an instrinsic Balmer decrement of $\Ha/\Hb=2.86$, consistent with Case B recombination at $T=10^4\mathrm{K}$ and $n_e=100\cmcub$ \citep{2003adu..book.....D,2006agna.book.....O}, and apply the \cite{1989ApJ...345..245C} extinction curve with $R_v = 3.1$. The results of this study are not dependent on our choice of a particular extinction law but are highly sensitive to the value of the Balmer decrement for each stacked bin. We ensure that the mean Balmer decrement of the individual low-z spectra that make up each bin are consistent with the Balmer decrement derived for their stacked spectra. We therefore assume that the extinction corrections applied to the stacked spectra are representative of the sample and not biased towards a subset of galaxies. Because of the limited number of \Hb\ detections in the high-z sample we are unable to make the same comparison between the average and stacked Balmer decrements. However, we note that the variation in Balmer decrements between different bins of high-z galaxies follows the correlation with SFR. The Balmer decrements of the stacked high-z bins vary from $\sim 3.8 - 5.1$, with higher values corresponding to bins containing galaxies with higher SFRs or stellar masses. The minimum value is consistent with the mean Balmer decrement of the subset of the high-z sample for which we measure both \Hb\ and \Ha\ at $\SN\geq 3$,  $F_{\Ha}/F_{\Hb} \sim 3.8$.  
        We conclude that the Balmer decrements of the stacked high-z bins are reasonable and take into account the larger uncertainties, compared to the stacked bins of low-z galaxies, when performing extinction corrections. 

        To estimate the errors on the extinction corrected emission-line fluxes we use a Monte-Carlo resampling technique. We perturb all five measured emission-line fluxes with the errors estimated from bootstrapping, and use the new Balmer decrement to correct the perturbed line fluxes. We repeat this process 1000 times to build a well-sampled distribution of corrected emission-line fluxes and take the standard deviation of the these corrected fluxes to be the errors on the emission-line fluxes for each bin. 
    

    \subsection{AGN contamination} 
        \label{sub:agn_contamination}

            We did not apply the \OIII/\Hb\ and \NII/\Ha\ ratios to investigate the AGN contamination for the majority of the high-z sample, which lacked detections of either \Hb, \OIII\ or \NII. We therefore investigate the extent of any remaining AGN contamination by comparing the position of our stacked samples to the theoretical maximum position of $z\sim 1.5$ star-forming galaxies on the BPT diagram from \citep{0004-637X-774-2-100} (dashed black line, Fig. \ref{fig:BPT}). Regardless of the property in which the samples are binned, the stacked data lie below the theoretical maximum position predicted at $z\sim 1.5$ predicted by \cite{0004-637X-774-2-100}. Moreover, the stacked samples are offset to lower \OIII/\Hb\ than the $z\sim 2.3$ sample of \cite{2014ApJ...795..165S} (orange line in Fig. \ref{fig:BPT}). 
    


\section{Deriving metallicities and ionisation parameters} 
  \label{sec:diagnosing}

    To investigate the evolution of the ionisation parameter we apply diagnostic methods that simultaneously constrain both the ionisation parameter and metallicity. We infer the metallicity in order to take into account its effect on the ionisation parameter and to account for the degeneracies between emission-line ratios and the metallicity and ionisation parameter. Although there exist numerous diagnostic methods which allow the metallicity to be inferred from the relative strengths of strong emission lines (see \citealt{2008ApJ...681.1183K} and \citealt{0067-0049-208-1-10} for discussions on metallicity diagnostics) the range of ionisation parameter diagnostics is far more limited. There currently exist three options for inferring the ionisation parameter, based on theoretical models of \HII\ regions. The first two options are the theoretically calibrated diagnostics of \cite{2002ApJS..142...35K} and \cite{2004ApJ...617..240K}. The third option is to use an algorithm that compares a user-supplied model grid of emission-line fluxes and ISM properties to the measured fluxes, e.g. \texttt{IZI} \citep{0004-637X-798-2-99} and \texttt{pyqz} \citep{2014ApJ...793..127V}. 

   We apply two diagnostic methods to infer ionisation parameters and metallicities. Both diagnostic methods rely on theoretical photoionisation models of idealised \HII\ regions, are sensitive to variations in metallicity and ionisation parameter and employ the five emission lines available for our high-z sample ($\OII\lambda 3727, \Hb, \OIII\lambda 5007, \Ha$ and $\NII\lambda 6584$). Our first method employs the IDL-based bayesian inference algorithm, \texttt{IZI}, developed by \cite{0004-637X-798-2-99}. Our second method makes use of the calibration of the $\ott$ and $\rtt$ ratios from \cite{2004ApJ...617..240K}, as described in \cite{2008ApJ...681.1183K}. We refer to the second diagnostic as KK04 diagnostic hereafter. 
         
    \subsection{IZI} 
      \label{sub:using_izi}

        We use \texttt{IZI} to infer the metallicity and ionisation parameter of the stacked galaxy samples based on the dust corrected emission-line fluxes and our theoretical photoionisation model grids. The formalism and statistical method employed by IZI are presented in detail in \cite{0004-637X-798-2-99}. By using \texttt{IZI}, we avoid some of the systematic uncertainties associated with the calibration of diagnostics based on a single strong emission line ratio \citep{0004-637X-647-1-244,0067-0049-167-2-177,2014ApJ...793..127V,0004-637X-798-2-99} and apply a model of our choice that is based on the most up-to-date atomic data and solar abundances. Moreover, most strong emission line diagnostics are unable to account for the asymmetry of the error bars on the derived parameters. Because \texttt{IZI} performs a full calculation of the joint Probability Distribution Function (PDF) of $Z$ and $\log(q)$, we are able to investigate the toplogy of the joint PDFs and account for any asymmetries. 

        We provide \texttt{IZI} with theoretical photoionisation models, based on multiple simulations of \HII\ regions using the \texttt{MAPPINGS V} photoionisation code, version 5.1.02. Our models are a subset of the ``pressure models'' described in detail in \cite{Kewley2017_inprep}. In brief, we use the \texttt{Starburst99} (\texttt{SB99}) models described in \cite{1538-3881-139-2-712} and \cite{2013ApJS..207...21N} and apply the stellar atmosphere models and tracks described in \cite{Kewley2017_inprep}. Our models are based on a Salpeter IMF \cite{1955ApJ...121..161S} with an upper mass limit of $100 \Msun$, but the choice of IMF has negligible impact on the optical emission-line fluxes used in our analysis. To account for the depletion of metals by dust as well as the impact on the cooling rate, we assume a base depletion of $[\mathrm{Fe}/\mathrm{H}]=-1.5\dex$ and apply the corresponding dust depletion model of \cite{2014arXiv1402.4765J}. Our simulations are based on continuous star formation models because these are more representative of the multiple sites of ongoing star-formation in star-forming galaxies than instantaneous bursts of star formation. We use an oxygen abundance scale and assume a bulk solar oxygen abundance of $\met=8.73$ consistent with the primordial solar abundance of \cite{2009ARA&amp;A..47..481A}. Our simulations are performed using a standard Maxwellian electron energy and assume a plane parallel geometry (as for the models used in the KK04 diagnostic).

        To account for the sensitivity between the ionisation parameter, metallicity and ISM pressure \citep[see ][]{0004-637X-647-1-244,0004-637X-774-2-100} we perform simulations that encompass a range of these parameters. Our simulated \HII\ regions have a detailed electron temperature and density structure that varies through the nebula based on the metallicity and the ionising radiation field. The ISM pressure in our simulated \HII\ regions ($P/k$) reflects the mechanical energy flux produced by the stellar population \citep[see also ][]{Kewley2017_inprep}.  The model metallicities are constrained by the stellar tracks used in the stellar population synthesis models.  Our simulated \HII\ regions span the following values of metallicity, ionisation parameter and pressure; 
        \begin{align*}
            12+\log(\OH) & = \{7.63, 8.23, 8.53, 8.93, 9.23 \} \\
            \begin{split}
             \log{\left(q \,/\cms \right)} & = \{6.50; 6.75; 7.00; 7.25; 7.50; 7.75; \\ 
                       & \qquad 8.00; 8.25; 8.50 \}        
           \end{split}
           \\
           \log([P/k] /\cmcub \, \mathrm{K}) & = \{5.0,5.5, 6.0,6.5,7.0,7.5, 8.0 \}  
        \end{align*} 

        It is beyond the scope of this work to simultaneously diagnose the ISM pressure as well as the metallicity and ionisation parameter. We therefore choose the appropriate ISM model pressure based on the relationships between the ISM pressure and the $\SII\lambda6717/\SII\lambda6731$ and $\OII\lambda3729/\OII\lambda3726$ ratios given in \cite{Kewley2017_inprep}. The relationship between the doublet ratios and ISM pressure depends upon the metallicity and ionisation parameter, as shown in Figures 9 and 13 of \cite{Kewley2017_inprep}. These relationships can be parameterised by functions of the doublet ratio, $\met$ and $\log(q)$. Based on the expected metallicity and ionisation parameter regimes $8.5<\met<8.9$ and $6.5<\log(q/\cms)<8.5$ (see results from KK04 diagnostic, figures \ref{fig:qz_M} to \ref{fig:qz_sSFR}), we predict the ISM pressures using the $\SII\lambda6717/\SII\lambda6731$ and $\OII\lambda3729/\OII\lambda3726$ ratios. The median $\SII\lambda6717/\SII\lambda6731$ ratios of the \Mstar-matched, SFR-matched, sSFR-matched and \Mstar-and-SFR-matched samples are $\sim 1.39, 1.29, 1.35$ and $1.29$ repectively, corresponding to ISM pressures of $P/k \sim 10^6, 10^{6.5}, 10^{6.5}$ and $ 10^{6.5} \cmcub \, \mathrm{K} $ \citep[see ][]{Kewley2017_inprep}. The median (and mean) $\OII\lambda3729/\OII\lambda3726$ ratio of the high-z sample is $\sim 1.2$ corresponding to an ISM pressure of $\sim 10^{6.5} \cmcub \, \mathrm{K} $. 

        We perform various tests before applying \texttt{IZI} to our stacked samples. Firstly, we ensure that when using the model fluxes as an input to \texttt{IZI} the inferred values of $Z$ and $\log(q)$ are those of the models. Secondly, we test the effect of increasing the errors and perturbing the input line fluxes by the errors. In order for the metallicity and ionisation parameter to be well constrained ($<0.3 \dex$), we require $\OII\lambda 3727$, \Hb, $\OIII\lambda 5007$, \Ha\ and $\NII\lambda 6584$ to be measured at $\SN \geq 5$. We also test the effects of using different combinations of strong emission-line fluxes. The choice of input line fluxes has a significant impact upon the derived values of $Z$ and $\log(q)$. The use of the $\SII\llambda 6717,6731$ flux, in addition to the other five lines, results in more accurate inferences of the metallicity (judged against the simulated \HII\ regions), especially in cases where the metallicity lies near the peak of the $\rtt$ vs $\met$ curve. The difference in inferred metallicities was up to 0.2dex. However, the addition of \SII\ had no noticeable impact on the ionisation parameter. The omission of \NII\ led to significant differences in metallicity (by up to $0.6\dex$ for individual galaxies) but had less effect on the ionisation parameter (up to $0.3\dex$). In contrast, the omission of \OIII\ or \OII\ greatly changed the measured ionisation parameter by $\delta \log(q) \sim 0.5\dex$.

        To allow for an unbiased comparison between the low- and high-z stacked samples we employ only the five emission lines measured for the $z\sim 1.5$ sample. We supply \texttt{IZI} with the dust-corrected emission-line fluxes and errors as well as the most suitable photoionisation model (based on the electron density). The error on our model emission-line fluxes is $0.05\dex$. We confirm that the inferred values of $Z$ and $\log(q)$ for each stacked sample are single-valued, bounded and represent a clear maximum in the joint 2D PDF. The uncertainties on $Z$ and $\log(q)$ represent the separation between the joint mode of $Z$ and $\log(q)$ and the $1\sigma$ confidence interval \citep[described in ][]{0004-637X-798-2-99}. In cases where the errors on $Z$ are more extended ($>0.3\dex$), the metallicity lies close to the peak of the $\rtt$ vs $\met$ curve.


        \begin{figure*}
          \begin{center}  
                \includegraphics[width=0.75\textwidth,trim={1.0cm 1.5cm 1.0cm 1.0cm},clip]{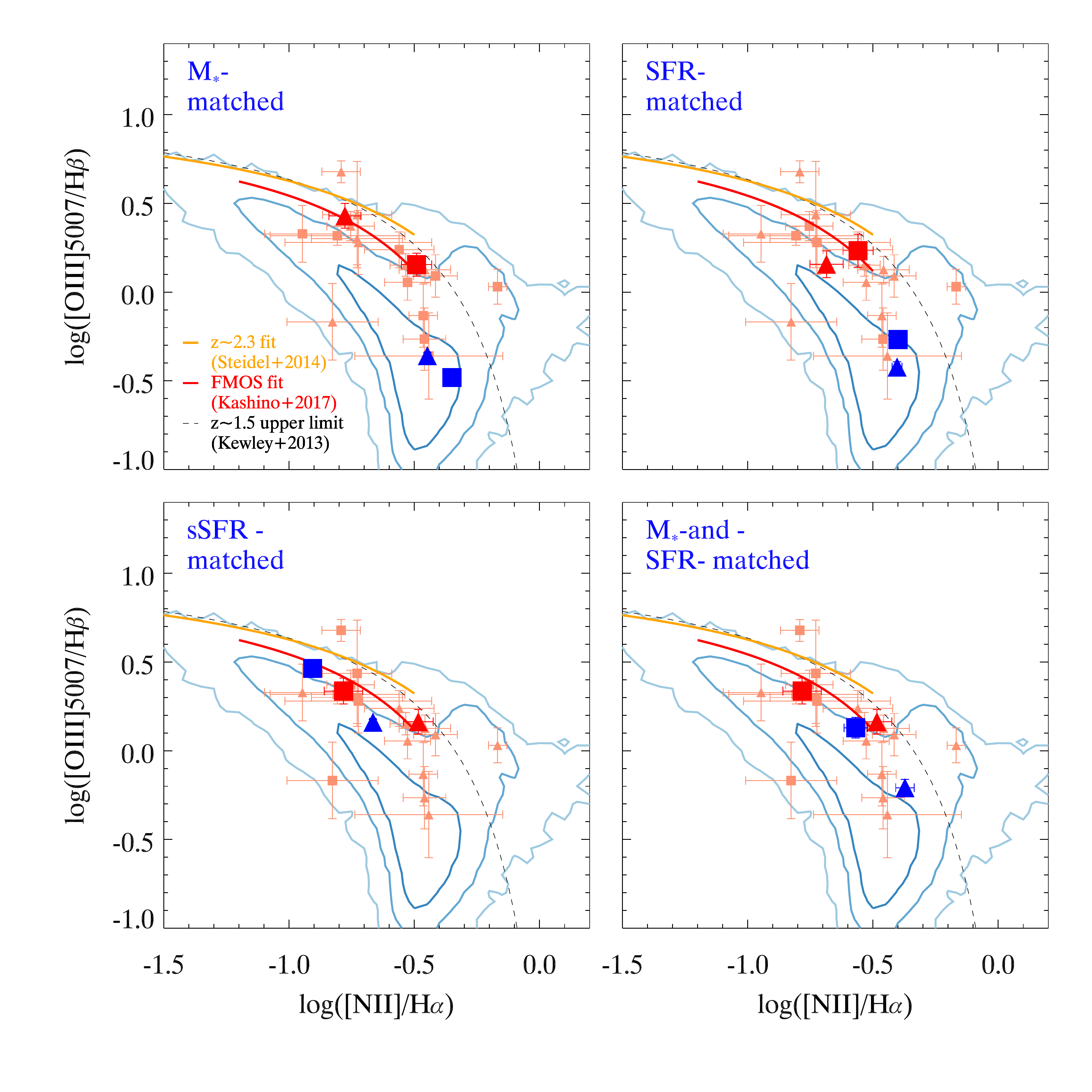} 
              \end{center}
            \caption{$\OIII\lambda 5007$/\Hb\ vs $\NII\lambda 6584$/\Ha\ for the stacked matched local (blue) and $z\sim 1.5$ (red) samples. Each panel shows the ratios derived from stacks of spectra binned in different quantities and matched  as labelled at the top left. The emission-line ratios of the stacked spectra (large filled symbols) are compared to the individual measurements of the high-z galaxies with $\Hb, \OIII\lambda5007,\Ha$ and $\NII\lambda6584$ detected at $\SN\geq3$ (small, light red filled symbols) and the distribution of the ``low-z star-forming catalogue'' ($1,2$ and $3\sigma$ limits represented by the light to dark blue contours). The symbols of the individual high-z galaxies indicate which bin they are assigned to, where triangles(squares) indicate galaxies assigned to the low (high) \Mstar, SFR or sSFR bins. The fits to $z\sim 2.3$ data from Steidel et al. (2014) (solid orange), the $z\sim 1.5$ FMOS sample of Kashino et al. (2017a) and the maximum theoretical ratios from Kewley et al. (2013) (solid black) are shown for comparison. Upper left: binned in \Mstar. Upper right: binned in SFR. Lower left: binned in sSFR. Lower right: \Mstar-and-SFR matched, binned in sSFR. }
            \label{fig:BPT}
        \end{figure*}  

         \begin{figure*}
          \begin{center}  
                \includegraphics[width=0.75\textwidth,trim={1.0cm 1.8cm 1.0cm 1.0cm},clip]{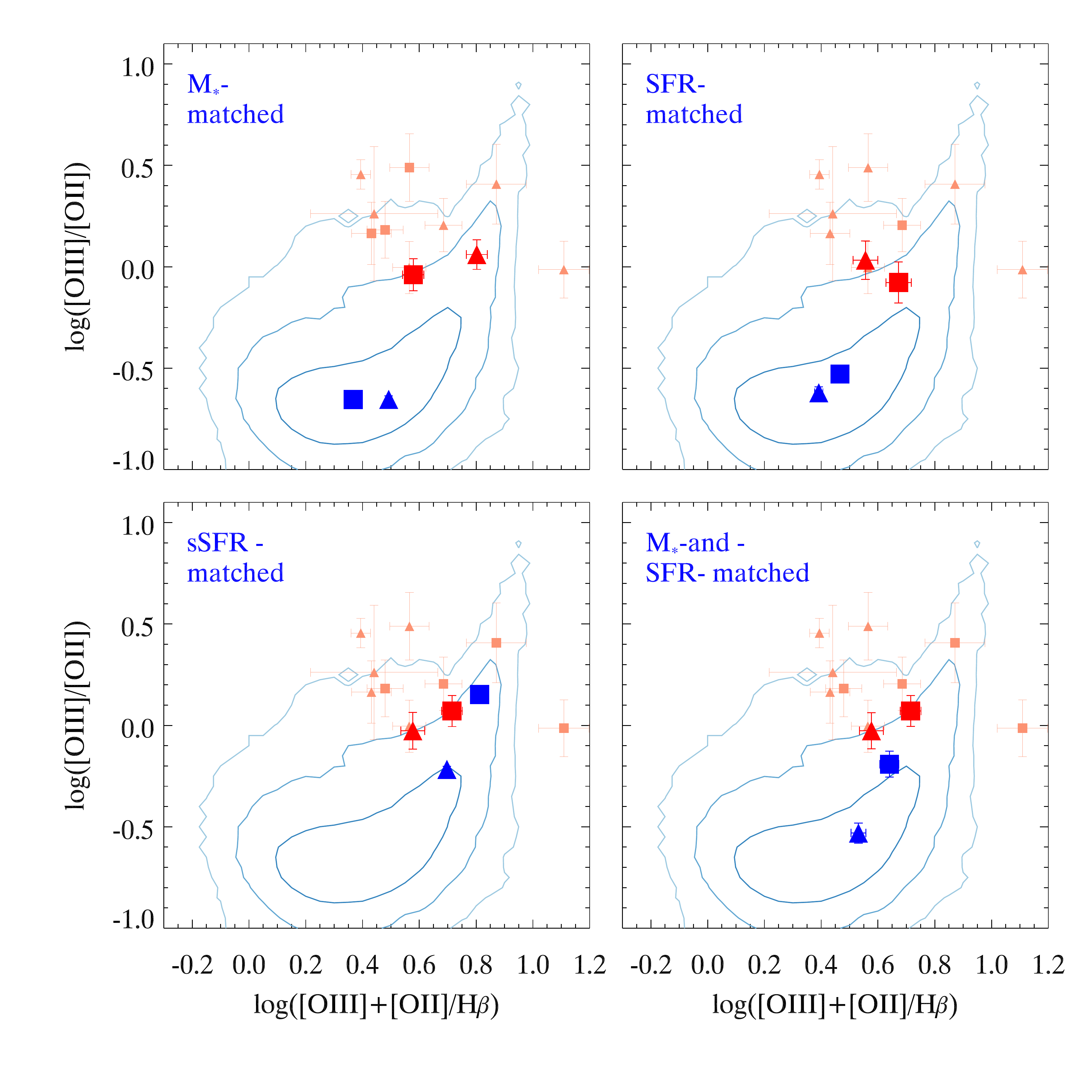} 
              \end{center}
            \caption{$\ott = \log(\OIII\llambda 4959,5007/\OII\llambda 3726,3729)$ vs $\rtt=\log(\OIII\llambda 4959,5007 + \OII\llambda 3726,3729/\Hb)$ for the four matched local (blue) and $z\sim 1.5$ (red) samples. Each panel shows different binned and stacked samples as labelled at the top left. The emission-line ratios of the stacked spectra (large filled symbols) are compared to the individual measurements of the high-z galaxies with $\OII,\Hb$ and $\OIII\lambda 5007$ detected at $\SN\geq3$ (small, light red filled symbols). The four panels, blue contours, and symbols are the same as in Fig. \ref{fig:BPT}}
            \label{fig:r23vs032}
        \end{figure*} 

         \begin{figure*}
          \begin{center}  
                \includegraphics[width=0.75\textwidth,trim={1.0cm 1.5cm 1.0cm 1.0cm},clip]{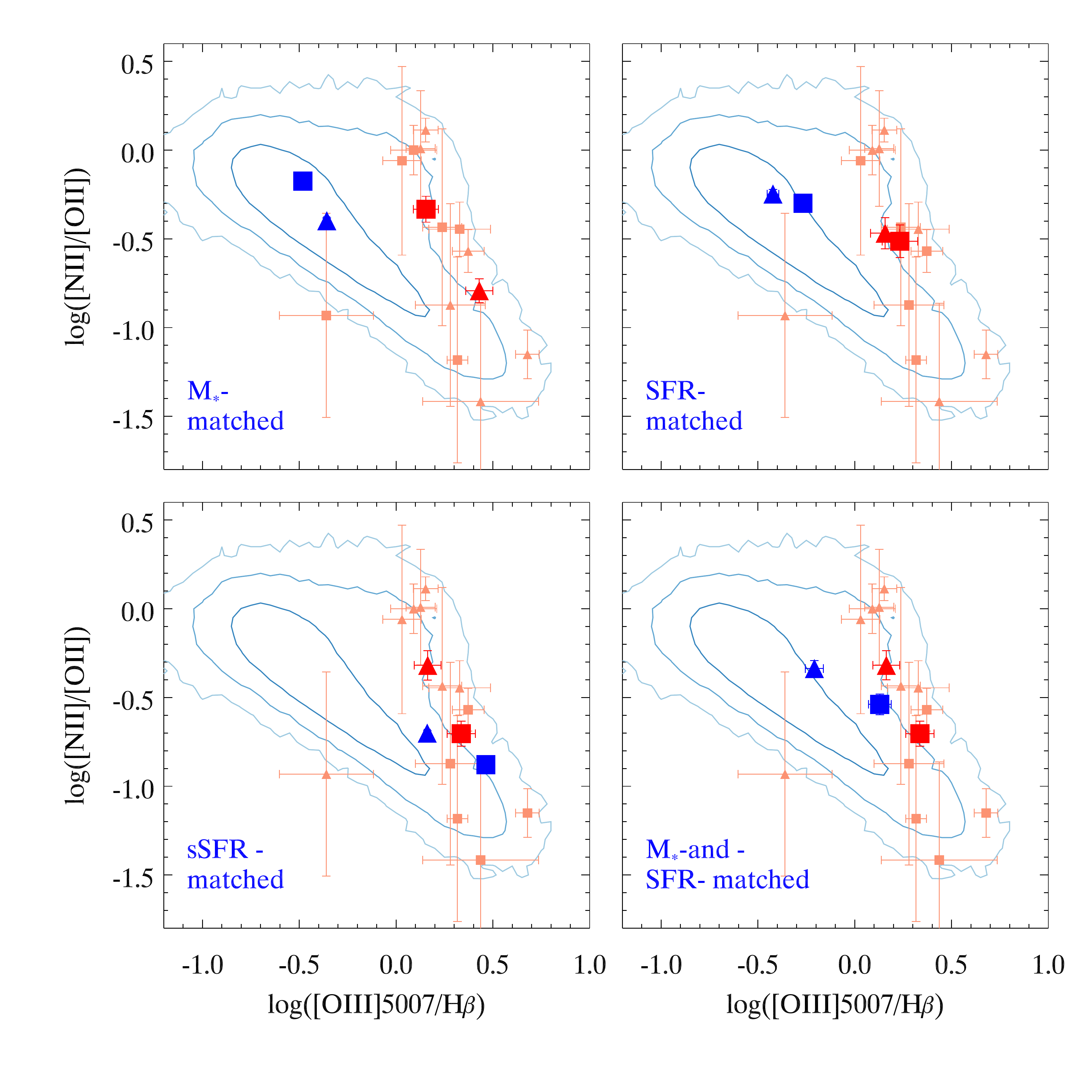} 
              \end{center}
            \caption{$\NII\lambda 6584/\OII\lambda 3727$ vs $\OIII\lambda 5007/\Hb$ for the four matched local (blue) and $z\sim 1.5$ (red) samples. Each panel shows different binned and stacked samples as labelled at the top left. The emission-line ratios of the stacked spectra (large filled symbols) are compared to the individual measurements of the high-z galaxies with $\OII,\Hb$ and $\OIII\lambda 5007$ and $\NII\lambda6584$ detected at $\SN\geq3$ (small, light red filled symbols). The four panels, blue contours, and symbols are the same as in Fig. \ref{fig:BPT}}
            \label{fig:n2o2vso3hb}
        \end{figure*} 

    \subsection{KK04 Diagnostic} 
        \label{sub:kk04_diagnostic}

        We also infer ionisation parameters and metallicities using the KK04 diagnostic \citep{2004ApJ...617..240K}. The KK04 diagnostic method is based on two line ratios; the ionisation parameter sensitive $\ott=\OIII\llambda 4959,5007/\OII\llambda 3726,3729$ ratio and metallicity sensitive $\rtt=\OIII\llambda 4959,5007+\OII\llambda 3726,3729/\Hb$ ratio. To derive the relationship between these line ratios and the values of $Z$ and $\log(q)$, \cite{2004ApJ...617..240K} applied the photoionisation grids of \cite{2002ApJS..142...35K}. These photoionisation models assume a single ISM pressure of $P/k = 10^5 \cmcub \, \mathrm{K} $. The models are described in detail in \cite{2002ApJS..142...35K}.

        Because the $\rtt$ and $\ott$ ratios are sensitive to both the metallicity and ionisation parameter, the iteration over both calibrations allows the metallicity and ionisation parameter to be jointly constrained. However, the $\rtt$ ratio is double valued with metallicity \citep[see ][]{2002ApJS..142...35K,0067-0049-167-2-177,0004-637X-647-1-244} and thus the KK04 method requires an initial metallicity guess to determine the branch over which to iterate. To determine the initial branch we use both the \NII/Ha\ and \NII/\OII\ ratios as described in Appendix of \cite{2008ApJ...681.1183K}. For each binned and stacked set of emission-line fluxes, the two \NII-based line ratios result in the same initial guess of metallicity branch, namely the upper branch \citep[$\met> 8.5$, ][]{2002ApJS..142...35K,2004ApJ...617..240K}. 

        The uncertainty on the values of $Z$ and $\log(q)$ inferred from the KK04 diagnostic tend to be smaller than the $1\sigma$ confidence intervals which result when using \texttt{IZI}. The comparitively smaller uncertainties are the result of a combination of factors. Firstly,  the metallicity branch is chosen before the KK04 diagnostic iterations are applied, and thus there is no uncertainty in the choice of the branch. Secondly, the unceratinties derived for the KK04 diagnostic are based solely on the errors on the emission-line fluxes and do not reflect the uncertainty of the models or the how well the data matches the models. 




       \begin{figure*}
            \begin{center}  
                \includegraphics[width=0.7\textwidth,trim={0.3cm 1.2cm 1.4cm 0.5cm},clip]{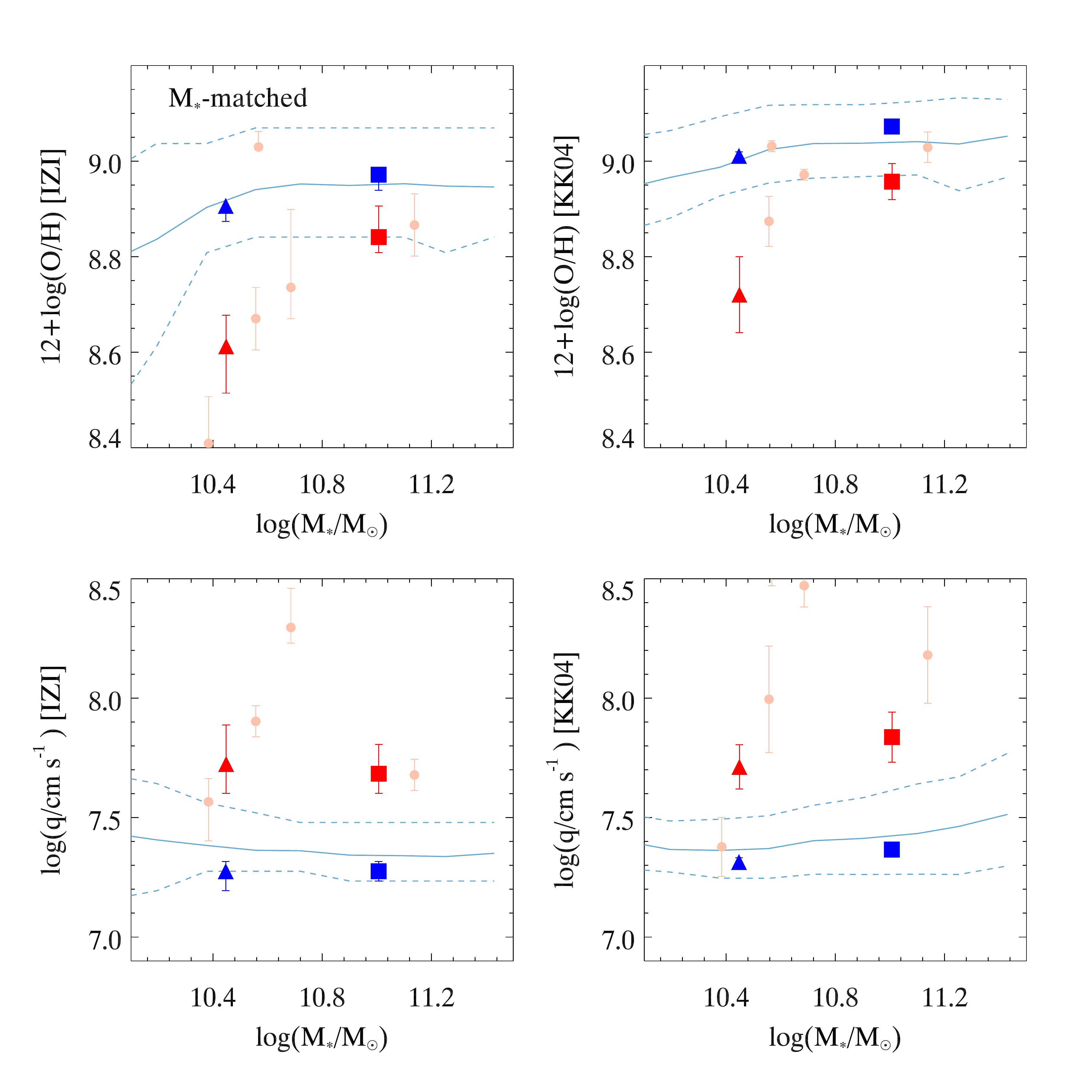} 
            \end{center}
            \caption{Metallicity and ionisation parameter as a function of \Mstar\ for the \Mstar-matched low-z (blue filled symbols) and high-z (red filled symbols) samples. The subset of $z~1.5$ galaxies for which $q$ and $Z$ could be inferred individually is shown for comparison (pink filled symbols). Left column: results from the IZI diagnostic. Right column: results based on the KK04 diagnostic. The symbols for the three bins are consistent with those in the top row of Fig. \ref{fig:match_global}. The mean metallicity and ionisation parameter (solid, light blue line), 16th and 84th percentiles (dashed light blue line) of the \Mstar-matched, SDSS sample derived via IZI (left panels) and the KK04 diagnostic (right panels) are shown as a function of \Mstar\ for comparison. }
            \label{fig:qz_M}
        \end{figure*}  

        \begin{figure*}
            \begin{center}  
                \includegraphics[width=0.7\textwidth,trim={0.3cm 1.2cm 1.4cm 0.5cm},clip]{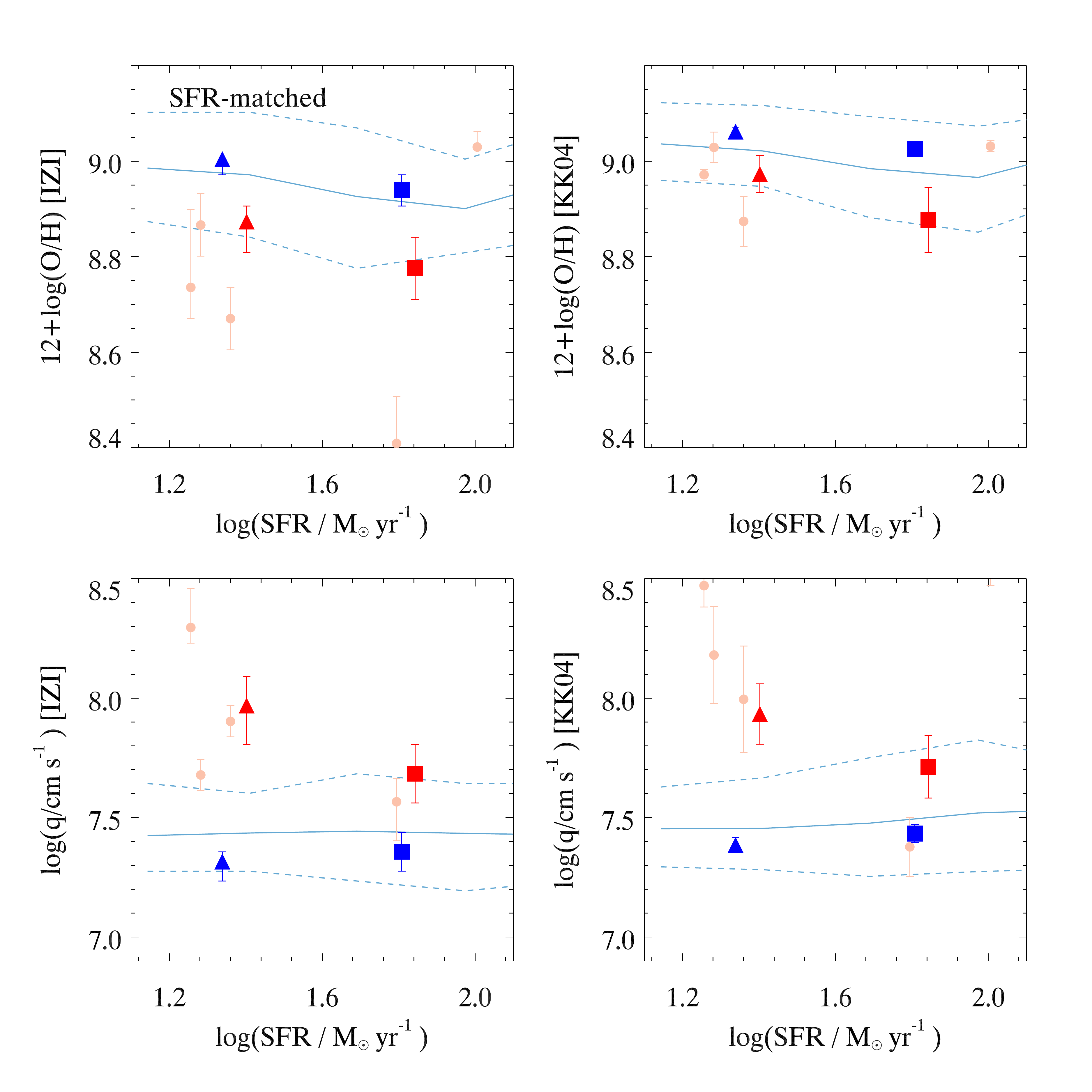}
            \end{center}
            \caption{Metallicity and ionisation parameter as a function of SFR for the SFR-matched low-z (blue filled symbols) and high-z (red filled symbols) samples. The subset of $z~1.5$ galaxies for which $q$ and $Z$ could be inferred individually is shown for comparison (pink filled symbols). Left column: results from the IZI diagnostic. Right column: results based on the KK04 diagnostic. The symbols for the binned and stacked samples are consistent with the second row in Fig. \ref{fig:match_global}. The mean metallicity and ionisation parameter (solid, light blue line), 16th and 84th percentiles (dashed light blue line) of the SFR-matched, SDSS sample derived via IZI (left panels) and the KK04 diagnostic (right panels) are shown as a function of SFR for comparison.}
            \label{fig:qz_SFR}
        \end{figure*} 
        \begin{figure*}
            \begin{center}  
                \includegraphics[width=0.7\textwidth,trim={0.3cm 1.2cm 1.4cm 0.5cm},clip]{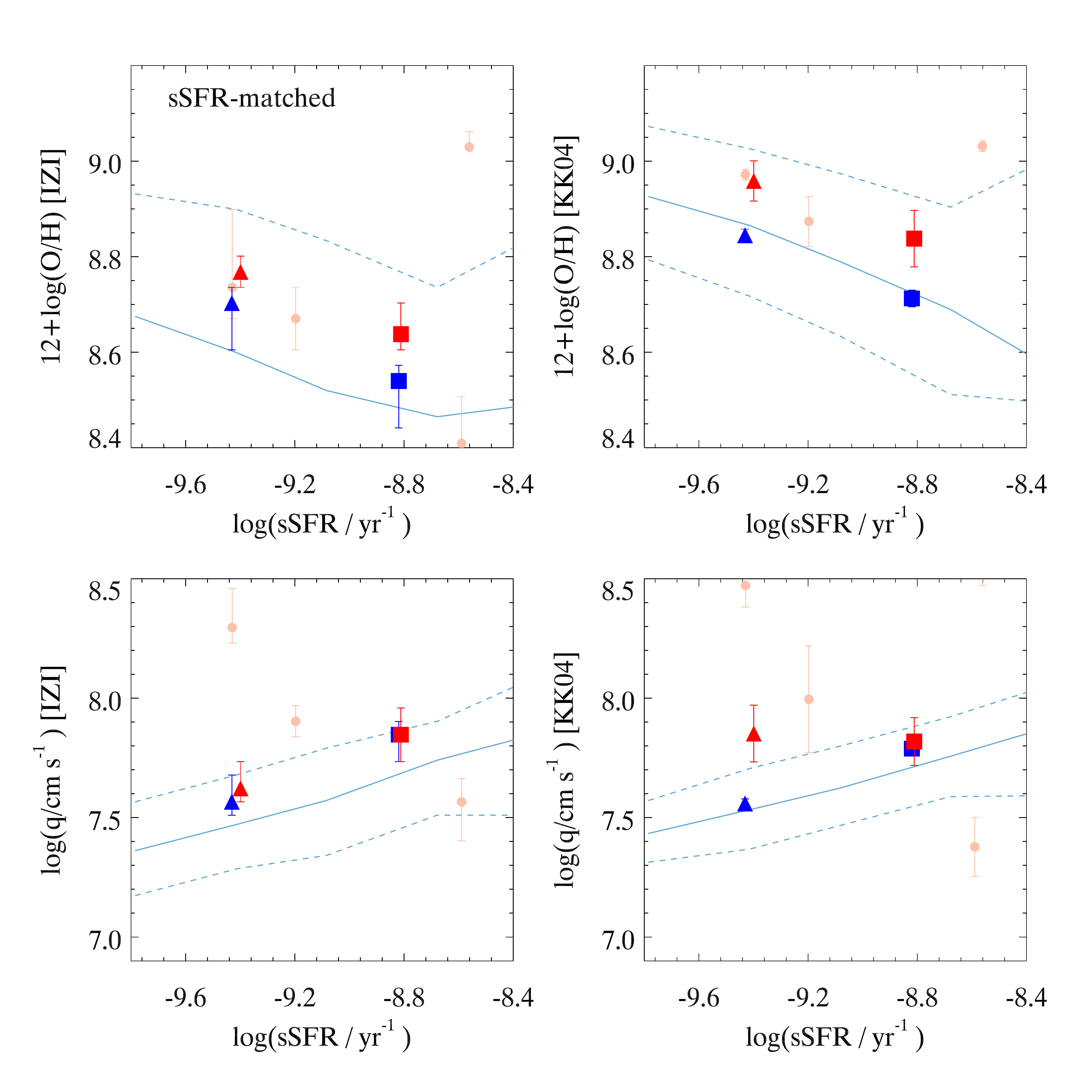} 
            \end{center}
            \caption{Metallicity and ionisation parameter as a function of sSFR for the sSFR-matched low-z (blue filled symbols) and high-z (red filled symbols) samples. The subset of $z~1.5$ galaxies for which $q$ and $Z$ could be inferred individually is shown for comparison (pink filled symbols). Left column: results from the IZI diagnostic. Right column: results based on the KK04 diagnostic. The symbols for the binned and stacked samples are the same as in the right panel of the third row in Fig. \ref{fig:match_global}. The mean metallicity and ionisation parameter (solid, light blue line), 16th and 84th percentiles (dashed light blue line) of the sSFR-matched, SDSS sample derived via IZI (left panels) and the KK04 diagnostic (right panels) are shown as a function of sSFR for comparison.}
            \label{fig:qz_sSFR}
        \end{figure*} 
        \begin{figure*}
            \begin{center}  
                \includegraphics[width=0.75\textwidth,trim={0.5cm 1.0cm 0.5cm 0.5cm},clip]{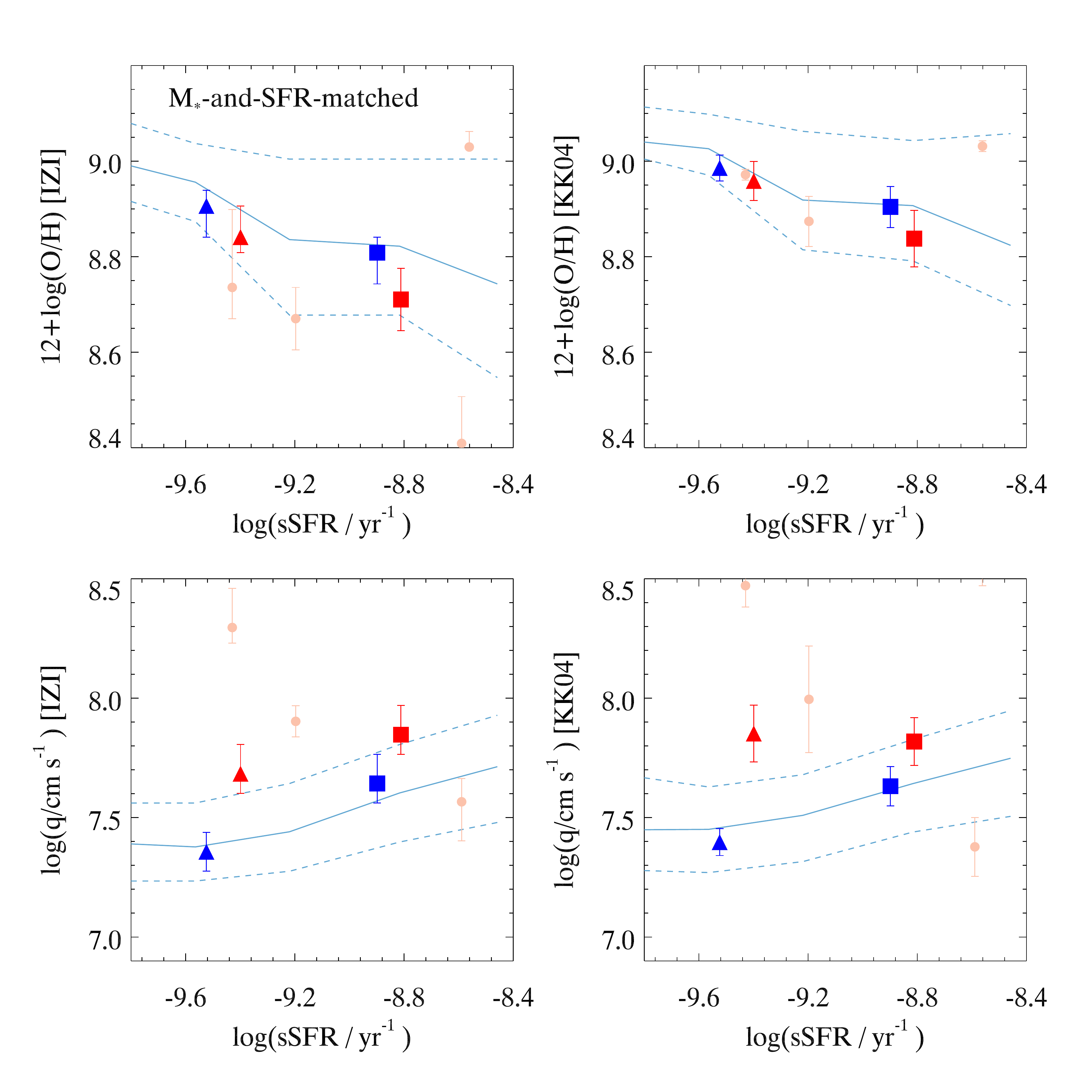} 
            \end{center}
            \caption{Metallicity and ionisation parameter as a function of sSFR for the \Mstar-and-SFR-matched low-z (blue) and high-z (red) samples. The subset of $z~1.5$ galaxies for which $q$ and $Z$ could be inferred individually is shown for comparison (pink filled symbols). Left column: results from the IZI diagnostic. Right column: results based on the KK04 diagnostic. Note that the position of the high-z sample is the same as in Fig. \ref{fig:qz_sSFR}. The symbols are the same as in the bottom row of Fig. \ref{fig:match_global}. The mean metallicity and ionisation parameter (solid, light blue line), 16th and 84th percentiles (dashed light blue line) of the \Mstar-and-SFR-matched, SDSS sample derived via IZI (left panels) and the KK04 diagnostic (right panels) are shown as a function of sSFR for comparison.}
            \label{fig:qz_MandSFR}
        \end{figure*} 


\section{Results} 
    \label{sec:results}

    \subsection{Emission-line ratios} 
        \label{sub:measured_emission_line_ratios}


        We compare the emission-line ratios of the matched low- and high-z stacked samples in Figures  \ref{fig:BPT} to \ref{fig:n2o2vso3hb}. We focus on three emission-line ratio diagrams; the \OIII/\Hb\ vs \NII/\Ha, or ``BPT'' diagram, \citep{1981PASP...93....5B}, the $\ott$ vs $\rtt$ diagnostic diagram and the \NII/\OII\ vs \OIII/\Hb\ diagram. The BPT diagram is commonly used to assess the dominant ionisation source of a galaxy, such as the presence of AGN. However, the position of galaxies on the BPT diagram is also sensitive to the metallicity, ionisation parameter and pressure \citep[discussed in detail in ][]{0004-637X-774-2-100}. The $\ott$ vs $\rtt$ diagnostic diagram is commonly used to trace the ionisation parameter and metallicity, because the $\ott$ ratio is highly sensitive to the ionisation parameter whereas the $\rtt$ is highly sensitive to metallicity \citep{2002ApJS..142...35K,2014MNRAS.442..900N,0004-637X-801-2-88,2016ApJ...816...23S}. However, the $\ott$ vs $\rtt$ diagram is not a clean diagnostic because the $\ott$ ratio is dependent on metallicity and the $\rtt$ diagnostic is dependent on the ionisation parameter \citep{2002ApJS..142...35K,Kewley2017_inprep}. The \NII/\OII\ vs \OIII/\Hb\ diagnostic diagram is mostly sensitive to the metallicity and ionising radiation source \citep[see.][]{0067-0049-208-1-10} but may also be a useful comparison of different galaxy samples \citep[e.g.][]{0004-637X-801-2-88}. \cite{0004-637X-801-2-88} argue that the an offset in \NII/\OII\ may indicate a difference in the N/O ratio at fixed metallicity. We do not interpret the \NII/\OII\ vs \OIII/\Hb\ diagram in the same manner. Instead, our work is based on the abundance scaling of N/O vs. O/H discussed in \cite{2017MNRAS.466.4403N}. 

        The consistency between the positions of matched low- and high-z samples on the \OIII/\Hb\ and \NII/\Ha\ diagram varies according to the property in which the samples are matched (red vs blue symbols in Fig. \ref{fig:BPT}).  For samples matched in \Mstar, the high-z bins exhibit  $\sim 0.7\dex$ larger \OIII/\Hb\ ratios and $0.2-0.3\dex$ smaller \NII/\Ha\ ratios than the equivalent low-z bins of \Mstar\ (top left panel). Similarly, for samples matched in SFR, the high-z bins exhibit $\sim 0.5\dex$ higher \OIII/\Hb\ and $0.1-0.3\dex$ lower \NII/\Ha\ ratios than the corresponding low-z bins (top right panel). In contrast, samples matched in sSFR exhibit consistent \OIII/\Hb\ ratios and show the opposite trend for \NII/\Ha\ (bottom left panel). The \NII/\Ha\ ratios of the high-z sSFR bins are  $\sim 0.1\dex$ higher than for the equivalent sSFR bin of low-z galaxies. For samples matched in both \Mstar\ and SFR the high-z bins exhibit $\sim 0.2 \dex$ larger \OIII/\Hb\ and $0.2\dex$ samller \NII/\Ha\ ratios than the matched low-z bins (bottom right panel). 

        The property in which the low- and high-z samples are matched also affects their relative positions on the $\ott$ vs $\rtt$ diagram (Fig. \ref{fig:r23vs032}). Both \Mstar\ bins of the high-z samples exhibit significantly higher $\ott$ ratios, $0.7-0.8\dex$, and $0.2-0.3\dex$ higher $\rtt$ ratios than the equivalent \Mstar\ bins of low-z galaxies (top left panel).  For samples matched in SFR, the high-z bins exhibit $\sim 0.6\dex$ larger $\ott$ and $0.2\dex$ larger $\rtt$ ratios than the equivalent low-z bins (top left panel). In contrast, the positions occupied by the the low- and high-z samples matched in sSFR are very similar, with the high-z bins of sSFR exhibit equivalent $\ott$ (within errors) and $<0.1\dex$ lower $\rtt$ ratios than the low-z bins. 

        As for the other emission-line ratio diagrams, we find that the positions occupied by the low- and high-z matched samples on the \NII/\OII\ vs \OIII/\Hb\ diagram are more closely matched for samples with equivalent sSFR (Fig. \ref{fig:n2o2vso3hb}). For the \Mstar-matched samples there is a greater offset between the \OIII/\Hb\ ratios of the low- and high-z sample than the \NII/\OII\ ratios ($\sim 0.6\dex$ vs $0.2-0.4\dex$). Similarly, for samples matched in SFR, the \OIII/\Hb\ ratios of the high-z sample are greater by $\sim 0.5\dex$ whereas the \NII/\OII\ ratios are smaller by $\sim 0.3\dex$. In constrast, the positions occupied by each bin of the sSFR-matched and \Mstar-and-SFR-matched low-z samples are consistent with the corresponding high-z bins, varying by $\leq 0.3\dex$ in either ratio (bottom row). The sSFR-matched and \Mstar-and-SFR-matched samples exhibit slight offsets in opposite ratios. The \NII/\OII\ ratios of the low- and high-z samples with equivalent sSFR is slightly offset, whereas the offset bewteen the \Mstar-and-SFR-matched samples is only in \OIII/\Hb. In addition, the positions occupied by  low-z sSFR-matched sample are consistent with individual galaxies in the high-z sample. 

        The differences in the \OIII/\Hb, $\ott$ and $\rtt$ ratios of the \Mstar-matched low- and high-z samples suggests that there is a significant evolution in the ionisation state of main-sequence galaxies from $z\sim 0.1$ to $z\sim 1.5$. However, once the sSFR is taken into account, the low- and high-z samples exhibit equivalent \OIII/\Hb\ and $\ott$. This suggests the evolution in sSFR is one of the main factors driving the observed offset of emission-line ratios at high redshift.     


    \subsection{Comparison with previous work} 
        \label{sub:comparison_with_previous_work}

        The differences between our low-z stacked samples, and between the two bins within each sample, are consistent with previous studies of the position of galaxies on the BPT and $\ott$ vs $\rtt$ diagrams. \cite{2016ApJ...828L..11D}, \cite{2008ApJ...678..758L} and \cite{2008MNRAS.385..769B} show that the position of galaxies on these diagrams is strongly correlated with the level of star formation activity, assessed via the sSFR, SFR surface density ($\Sigma_\mathrm{SFR}$) or excess EW(\Ha).  Galaxies with higher levels of star-formation activity relative to typical star-forming galaxies of the same stellar mass (and at the same epoch) tend to lie above the local star-forming abundance sequence on the BPT diagram \citep{2008MNRAS.385..769B,2008ApJ...678..758L}, exhibiting both higher \OIII/\Hb\ \citep{2016ApJ...828L..11D} and $\ott$ ratios \citep{2014MNRAS.442..900N,2016ApJ...816...23S} than the more typical low-z, star-forming galaxies. Likewise, we find that the low-z, sSFR-matched sample exhibits significantly higher \OIII/\Hb, $\ott$ and $\rtt$ ratios than either bin of the \Mstar-matched sample, which has a lower mean sSFR (blue filled symbols in top vs bottom left panels of Figures \ref{fig:BPT} and \ref{fig:r23vs032}). Moreover, the high sSFR bin of the sSFR-matched sample exhibits higher \OIII/\Hb\, $\ott$ and $\rtt$ than the low sSFR bin (triangle vs square in bottom left panel of Figures \ref{fig:BPT} and \ref{fig:r23vs032}). 

        The position of our high-z sample on the diagnostic diagrams is also consistent with previous theoretical and high-z observational studies \citep[e.g,.][]{0004-637X-774-2-100,2014ApJ...785..153M,2016ApJ...816...23S,2016ApJ...828L..11D,2017ApJ...835...88K}. Our high-z sample (both individual and stacked) exhibits higher \OIII/\Hb\ ratios, higher $\ott$ and lower \NII/\Ha\ than the majority of low-z, star-forming galaxies (blue contours in Figures \ref{fig:BPT} and \ref{fig:r23vs032}) and the \Mstar-matched sample. Assuming that the star-forming abundance sequence (on the BPT diagram) increases with redshift, as suggested by \cite{2013ApJ...774L..10K}, then we would also expect our sample to lie between the local sequence and the $z\sim 2.3$ sample of \cite{2014ApJ...795..165S}. We find that our sample of high-z galaxies indeed lies below the $z\sim 2.3$ sample of \cite{2014ApJ...795..165S} (see orange line in Fig. \ref{fig:BPT}).  Moreover, our high-z sample occupies comparable positions on the $\ott$ vs $\rtt$ and \NII/\OII\ vs \OIII/\Hb\ diagrams to the $z\sim 2.3$ MOSDEF sample presented in \cite{0004-637X-801-2-88} and \cite{2016ApJ...816...23S}, although our individual detections exhibit greater scatter and lower S/N.


    \subsection{Metallicities} 
        \label{sub:metallicity}

        We compare the metallicity of each matched bin of low- and high-z galaxies in the top rows of Figures \ref{fig:qz_M} to \ref{fig:qz_MandSFR}. Each matched low- and high-z bin is represented by the same symbol, with the low-z sample in blue and the high-z bin in red. For comparison, we show the average metallicity of the \Mstar-, SFR-, sSFR-, and \Mstar-and-SFR-matched, low-z samples as a function of the matched property (blue solid and dashed lines). The individual metallicities of each low-z matched sample are derived via the same methods as for the stacked bins. We also show the subsample of high-z galaxies for which we can individually diagnose the metallicity and ionisation parameter.

        The offset between the metallicities of the low- and high-z samples varies according to the property in which they are matched. The \Mstar-matched, low-z bins exhibit $0.15-0.3\dex$ higher metallicities than the matched high-z bins and the SFR-matched, low-z bins exhibit $0.15\dex$ higher metallicities than the equivalent high-z bins. In contrast, the sSFR-matched low-z bins exhibit $0.1-0.2\dex$ lower metallicties than the corresponding high-z bins. For the \Mstar-and-SFR-matched samples, the low- and high-z bins have metallicities that are consistent within the errors. 
 
        The metallicity of the \Mstar-matched samples increases with \Mstar, consistent with previous studies of the MZ relation \citep[e.g.][]{2004ApJ...613..898T,2013ApJ...771L..19Z,0004-637X-792-1-75,2017ApJ...835...88K}. For both the low- and high-z stacked samples the higher stellar mass bin exhibits a higher metallicity than the low mass bin. Likewise, the average metallicity of the \Mstar-matched sample and full low-z star-forming catalogue increases as a function of metallicity, consistent with the MZ relation of \cite{2013ApJ...771L..19Z}. We also find a significant offset between the metallicity of the low- and high-z matched samples, consistent with the evolution of the MZ relation observed by \citep{2013ApJ...771L..19Z}. 
        
        The difference between the metallicities of the matched low- and high-z bins is independent of the diagnostic method used. However, the exact values of metallicity are offset between the diagnostics, with the KK04 diagnostic predicting $\sim 0.1\dex$ higher metallicities than \texttt{IZI} for each stacked bin. There are three main reasons for the discrepancy between the diagnostics; the models on which the diagnostics are based, the use of the measured emission lines and the choice of grid pressure.

        We test the effects of the difference between the models via the two diagnostics using a sample of \HII\ regions from \cite{1998AJ....116.2805V}. To evaluate the difference between the models we restrict \texttt{IZI} to the \OII, \Hb\ and \OIII\ lines, used by the KK04 diagnostic. We apply the $\log{P/k}=5$ pressure grid, the same ISM pressure used for the KK04 diagnostic, and use only the 136 (out of 188) \HII\ regions which lie on the upper metallicity branch $\met>8.3$. To ensure that \texttt{IZI} chooses the correct branch we apply a step function prior on the metallicity \citep[see also][]{0004-637X-798-2-99}.  Based on the same set of lines, the KK04 diagnostic predicts $0.13\dex$ higher metallicities, on average, than \texttt{IZI}, with a dispersion of $0.03\dex$ about the mean offset. The difference between the two diagnostics when using the same lines is due to the stellar evolution and photoionization model revisions that have occurred between 2002 and 2017, which include the introduction of metal opacities in the stellar atmospheres, updated atomic data and abundance sets in the photoionization models, and a more detailed treatment of dust physics during radiative transfer calculations.
        
        The second difference between the diagnostics applied to the stacked samples is the treatment of the \NII\ and \Ha\ lines. The KK04 diagnostic gives all the weight to the relative strengths of \OII, \Hb\ and \OIII\ via the $\ott$ and $\rtt$ ratios, whereas we also account for the effect of \NII/\OII\ and \NII/\Ha\ when using \texttt{IZI}. Both the \NII/\OII\ and \NII/\Ha\ ratios are highly sensitive to the metallicity and therefore allow for more accurate inferences of the metallicity \citep[e.g.][]{0004-637X-798-2-99,0067-0049-208-1-10}. The addition of the \NII\ flux as an input to \texttt{IZI} when diagnosing the metallicities of the 136 \ion{H}{II} regions (on the upper metallicity branch) from \cite{1998AJ....116.2805V} resulted in an increased metallicity offset, of $0.24\dex$, with a dispersion of $0.13\dex$. We note that the additional use of the \SII\ flux resulted in a similar metallicity offest of $0.23\dex$, with a dispersion of $0.13\dex$ about the mean offset. 
    

    \subsection{Ionisation parameters} 
        \label{sub:metallicity_and_ionisation_parameter}

        We compare the ionisation parameters of each matched bin of low- and high-z galaxies in the bottom rows of Figures \ref{fig:qz_M} to \ref{fig:qz_MandSFR}. Each matched low- and high-z bin is represented by the same symbol, with the low-z sample in blue and the high-z bin in red. For comparison, we show the average ionisation parameter of the \Mstar-, SFR-, sSFR-, and \Mstar-and-SFR-matched, low-z samples as a function of \Mstar, SFR, sSFR and sSFR respectively (blue solid and dashed lines) derived via the same methods as for the stacked bins. We also show the subsample of high-z galaxies with high enough $\SN$ in all five lines to individually diagnose the metallicity and ionisation parameter.

        The ionisation parameters of the low- and high-z samples are only consistent for samples matched in sSFR.  We recover an offset of 0.3-0.5\dex\ in the ionisation parameter between the high-z sample and \Mstar-matched low-z sample (Fig. \ref{fig:qz_M}). Thus, we show that there is an evolution of $\sim 0.4\dex$ in the ionisation parameter of main-sequence galaxies from $z\sim 0.1$ to $z\sim 1.5$. We recover a similar offset, between the high-z sample and SFR-matched low-z sample, although the highest SFR bin of the SFR-matched samples is more closely matched than the lowest SFR bin. The ionisation parameters of low- and high-z bins matched in both \Mstar\ and SFR are more closely matched than for either the \Mstar- or SFR-matched samples (bottom row, Fig. \ref{fig:qz_MandSFR}). We note that the sSFRs of the low-z \Mstar-and-SFR-matched low-z sample are slightly lower than for the same bins of the high-z sample (as shown by the difference in sSFR in Fig. \ref{fig:qz_MandSFR}). In contrast to the other matched samples, the ionisation parameters of the sSFR-matched low- and high-z matched samples are consistent within 1$\sigma$ uncertainties (bottom panels, Fig. \ref{fig:qz_sSFR}).  

        The difference between the ionisation parameters of the matched samples is mostly independent of the diagnostic method used. For most of the matched samples the two different diagnostics give the same offset between the ionisation parameter and metallicity of the low vs high-z bins. We find only one instance for the stacked samples where the ionisation parameter predicted by the KK04 diagnostic appears significantly greater than that inferred from \texttt{IZI}, the lowest sSFR bin of the high-z sample (Figures \ref{fig:qz_sSFR} and \ref{fig:qz_MandSFR}). We note that the discrepancy is likely to result from the difference in assumed ISM pressures. The KK04 diagnostic, which is based on a lower pressure, is ascribing the emission-line properties to a higher ionisation parameter than \texttt{IZI}, which uses a higher model pressure. 

        We find differences of up to $0.17\dex$ between the ionisation parameters inferred for the stacked samples via the two diagnostics. However, there does not appear to exist a systematic offset for the stacked samples. For the \Mstar-matched and SFR-matched, low-z samples the ionisation parameters inferred via the KK04 diagnostic are $\sim 0.05\dex$ higher than inferred via \texttt{IZI} whereas for the sSFR-matched, low-z sample the ionisation parameters inferred via the KK04 diagnostic are $0.05\dex$ lower. 

        We test the difference between the ionisation parameters inferred via the two diagnostics using the 136 \HII\ regions from \cite{1998AJ....116.2805V} with which we also investigate differences in the metallicity. We use the $\log{P/k}=5$ pressure grid and applying the same set of emission lines used in the KK04 diagnostic when using \texttt{IZI}. Applying the same criteria, we find that the KK04 diagnostic predicts $0.07\dex$ higher ionisation parameters, on average, than \texttt{IZI} with a dispersion about the mean of $0.08\dex$, which is within the errors of the models ($0.1\dex$). If we instead use the set of five lines applied to our stacked samples we find a mean offset of $0.09\dex$ with a $0.1\dex$ dispersion. The additional use of the total $\SII\llambda 6717,6713$ flux resulted in a greater mean offset of $0.14\dex$ with a dispersion about the mean of $0.1\dex$. 
    



\section{Discussion} 
    \label{sec:discussion}

        \begin{figure*}
              \begin{center}  
                    \includegraphics[width=0.7\textwidth,trim={0.0cm 0.0cm 0.5cm 0.5cm},clip]{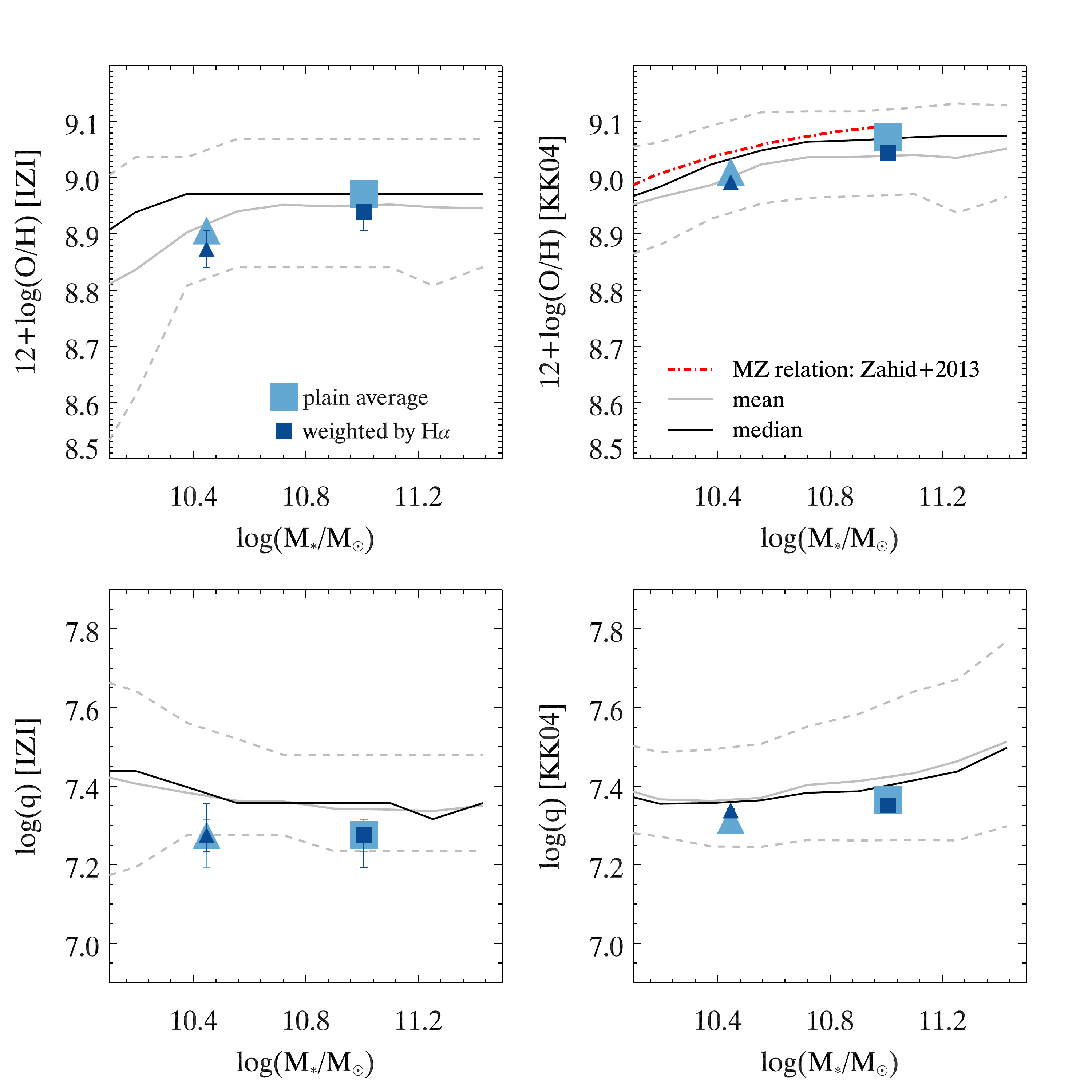} 
              \end{center}
            \caption{ Metallicity (top row) and ionisation parameter (bottom row) as a function of \Mstar\ for the low-z, \Mstar-matched sample. The mean and median of the sample, in bins if \Mstar, are shown by the grey and black lines respectively. The dashed grey lines indicate the 16th and 84th percentiles. The results from two stacking methods are compared; the simple average method we have chosen (large light blue symbols) and weighting by the \Ha\ flux prior to stacking (small dark blue symbols).  Left column: results based on the IZI diagnostic. Right column: results based on the KK04 diagnostic. The MZ relation of Zahid et al. (2013), based on the KK04 diagnostic, is shown for comparison (red dashed line, upper right panel).}
            \label{fig:test_mz_stack}
        \end{figure*} 

    \subsection{Validation of methods} 
        \label{sub:validity}

        We investigate how stacking spectra impacts the emission-line and ISM properties we derive for each sample. To assess the impact of stacking the high-z spectra we compare the emission-line ratios derived via stacking to the individual ratios of the parent samples (Figures \ref{fig:BPT} to \ref{fig:n2o2vso3hb}).  Moreover, the stacked high-z bins exhibit \OIII/\Hb\ and \NII/\Ha\ ratios that are consistent with the fit to the larger FMOS sample presented in \cite{2017ApJ...835...88K} (see solid red line in Fig. \ref{fig:BPT}). 
         
        We investigate the validity of our stacking method via the relationship between metallicity/ionisation parameter and \Mstar\ in Fig. \ref{fig:test_mz_stack}. The metallicities and ionisation parameters inferred based on the stacked spectra are consistent with both the mean (solid grey) and median (solid black) values of the individual sample. Moreover, the relationship between the metallicity and \Mstar\ of our \Mstar-matched low-z sample is consistent with the MZ relation derived by \cite{2013ApJ...771L..19Z} (Fig. \ref{fig:test_mz_stack}). The median metallicities of the \Mstar-matched sample (solid black line in the upper right panel of Fig. \ref{fig:test_mz_stack}) are slightly lower than the MZ relation of \cite{2013ApJ...771L..19Z}, especially at the high mass end. However, the MZ relation of \cite{2013ApJ...771L..19Z} is based on a lower stellar mass range, indicated by the red dashed line in Fig. \ref{fig:test_mz_stack}. The differences between our MZ relation and that of \cite{2013ApJ...771L..19Z} are due to the difference in AGN selection criteria (\cite{2001ApJ...556..121K} vs \cite{2003MNRAS.346.1055K}), the different stellar mass ranges and the smaller size of our sample.

        We further assess whether our stacking method results in any bias in the metallicity or ionisation parameter within the low-z (SDSS) sample.  As suggested by \cite{2017ApJ...835...88K}, simply taking the mean spectrum may yield emission line measurements that are weighted towards more luminous and therefore lower metallicity galaxies. Moreover \cite{2016ApJ...827...35T} show that galaxies with higher sSFRs may have a larger fraction of light covered by the SDSS fibre and may therefore inlcude lower metallicity regions of the galaxy. To test the extent to which our stacked emission line measurements are weighted towards a more luminous subsample we scale the spectra of the low-z, \Mstar-matched sample by the inverse of the peak \Ha\ flux prior to stacking. We find no significant difference between the Balmer decrements, metallicity or ionisation parameter derived via the simple averaging method vs. the \Ha\ scaled method (Fig. \ref{fig:test_mz_stack}). We therefore conclude that our stacking method does not significantly bias our results to either more luminous or low metallicity galaxies. We also argue that the variation in aperture covering fractions with sSFR do not bias our stacked results as the metallicities and ionisation parameters inferred based on the two stacked bins of sSFR-matched low-z galaxies are consistent with the mean values inferred individually for the sample (blue line vs filled blue symbols in Fig. \ref{fig:qz_sSFR}). 

        We note that the ISM properties discussed in this work represent global averages, which do not account for the dispersion amongst galaxies. Our emission-line spectra represent the luminosity-weighted integrated spectra of regions of ionised interstellar gas which fall within our aperture. We further smooth the effects by stacking these integrated spectra in bins of global galaxy properties. We thereby assume that the metallicities and ionisation parameters derived for the galaxy samples are representative of the conditions within the most luminous star-forming regions. Moreover, our work does not attempt to take into account the effects of diffuse gas emission discussed in \cite{2017ApJ...850..136S}. Instead, the impact of diffuse gas on the diagnostic methods used here will be discussed in \cite{Poetrodjojo}. 

        We have modelled the emission from the galaxies within our sample as coming from ``ionisation-bounded'' \ion{H}{II} regions, for which the size is determined by the ionisation equilibrium between the production rate of hydrogen-ionising photons and the hydrogen recombination rate. Outside the ``ionisation-bounded'' \ion{H}{II} region the hydrogen gas is 99\% neutral. In the alternative, ``density-bounded'', scenario, the size of the \ion{H}{II} region is defined by the distance at which the ionising photons run out of matter to ionise. For a ``density-bounded'' \ion{H}{II} region, the gas density is low enough that the stars may completely ionise the hydrogen in the gas cloud, leaving no neutral HI gas outside the \ion{H}{II} region. Density-bounded \ion{H}{II} regions may have significantly smaller \NII\ and \OII\ zones, resulting in larger \OIII/\Hb\ and \OIII/\OII\ ratios than what would be observed for ionisation-bounded \HII\ regions \citep{2008MNRAS.385..769B,0004-637X-774-2-100,2014MNRAS.442..900N}. If our simulations are applied to density-bounded nebulae, the larger line ratios may be interpreted as greater ionisation parameters. Both ionisation- and density-bounded nebulae have been observed in the local group \citep{0004-637X-755-1-40}. However, it is still unclear which scenario is most typical in main-sequence star-forming galaxies, either locally or at high redshift. 
    

    \subsection{The evolution of ionisation parameter} 
        \label{sub:evolution_of_ionisation_parameter}

        Our results indicate that the ionisation parameter evolves with the sSFR of galaxies. We find that there is an evolution in the ionisation parameter for main-sequence galaxies, from $z\sim 0.1$ to $z\sim 1.5$. However, once the differences in sSFR (from the evolution of the main sequence) are accounted for we find little evolution in the ionisation parameter. Samples with consistent, high sSFRs exhibit equivalent, average ionisation parameters. In contrast, samples of low-z galaxies for which only the \Mstar\ or SFR is equivalent to our high-z sample, exhibit significantly lower average ionisation parameters. Moreover, both the sSFR- and \Mstar-and-sSFR matched low-z samples appear to exhibit a positive correlation between the average ionisation parameter and sSFR (i.e. solid blue lines in Figures \ref{fig:qz_sSFR} and \ref{fig:qz_MandSFR}). 

        \begin{figure*}
            \begin{center}  
            \includegraphics[width=0.95\textwidth,trim={0.5cm 0cm 0.5cm 0.0cm},clip]{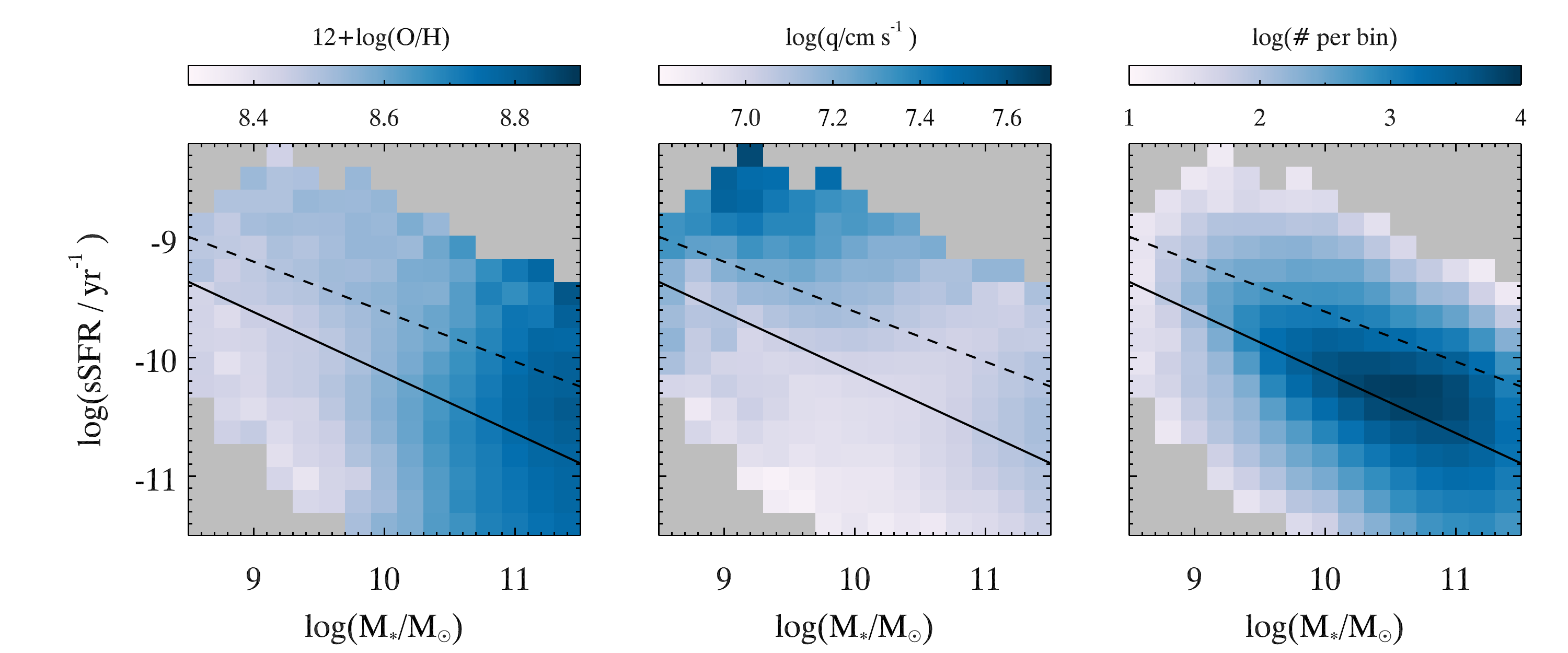} 
            \end{center}
            \caption{ Mean metallicity (left) and ionisation parameter (middle) as a function of \Mstar\ and sSFR for the full low-z star-forming catalogue. Each bin of \Mstar\ and sSFR contains at least 20 galaxies. The number of galaxies per bin is shown in the right hand panel. The metallicity and ionisation parameter are based on \texttt{IZI}, using \OII, \Hb, \OIII, \Ha\ and \NII. The main-sequence fits from Speagle et al. (2014) at $z\sim 0.1$ and $z\sim 0.3$ are shown by the solid and dashed black lines respectively.}
            \label{fig:qz_global}
        \end{figure*}

        To validate the results of the stacked analysis and investigate the dependence of the ionisation parameter on metallicity, \Mstar\ and sSFR we determine the metallicity and ionisation parameter individually for each galaxy in the low-z star-forming catalogue.  We derive the metallicity and ionisation parameters via \texttt{IZI} using the strong emission lines, $\OII\llambda 3726,3729$, \Hb, $\OIII \lambda 5007$, \Ha, $\NII\lambda 6584$ and $\SII\llambda 6717,6731$, and photoionisation grids with an ISM pressure selected according to the $\SII\lambda6717/\SII\lambda6731$ ratio, as described in Section \ref{sub:using_izi}. We separate the low-z star-forming catalogue into bins of \Mstar\ and sSFR, each containing at least 20 galaxies, and show the mean metallicity and mean ionisation parameter inferred via \texttt{IZI} as well as the number of galaxies per bin in Fig. \ref{fig:qz_global}. In each panel, we show the main sequence at $z\sim 0.1$ and $z\sim 0.3$ from \cite{2014ApJS..214...15S} (solid and dashed black lines respectively). We note that the main trends are the same when using the KK04 diagnostic. However, the KK04 diagnostic shows a greater sensitivity of metallicity to sSFR than when we apply \texttt{IZI} and account for variations in pressure. 

        We find that the ionisation parameter shows a strong variation with sSFR for the $z<0.3$ sample of star-forming galaxies (Fig. \ref{fig:qz_global}), confirming the results from the matched samples. Within the $z<0.3$ sample, the ionisation parameter scales most strongly with the sSFR, and only shows a weak dependence on \Mstar, in that the highest ionisation parameters are also found at the lowest values of \Mstar. In contrast to the ionisation parameter, the metallicity scales most strongly with \Mstar, rather than sSFR (although the strength of the correlation with sSFR is sensitive to the diagnostic as well as the treatment of dust and aperture effects \citep{2016ApJ...827...35T}.  Along the main sequence (black solid line), the mean metallicity increases significantly (by $\sim 0.6\dex$) with \Mstar, whereas the ionisation parameter shows no significant correlation with \Mstar.

        Both our stacked samples and the low-z star-forming catalogue indicate that high ionisation parameters are not driven simply by low metallicities. Although the sSFR-matched, low-z sample has lower metallicities than the high-z sample, the ionisation parameters of the two bins are consistent (see Figure \ref{fig:qz_sSFR}). Within the low-z star-forming catalogue the metallicity increases with \Mstar\ for $ \sSFR < 10^{-9.5} \peryr $, whereas the ionisation parameter shows little variation. Bins with the lowest metallicities do not correspond to the highest ionisation parameters. Nor does the mean ionisation parameter follow the trend of increasing metallicity along the main sequence. 

        The dependence of the ionisation parameter on sSFR, shown by our matched samples and within the low-z star-forming catalogue, is consistent with the correlations presented in \cite{2014MNRAS.442..900N,2015ApJ...812L..20K,2016ApJ...816...23S} and \cite{2016ApJ...822...62B}. \cite{2016ApJ...822...62B} compare two sets of local SDSS galaxies, selected according to their position on the BPT diagram. Those occupying positions consistent with main-sequence galaxies at $z\sim 2.3$, the ``local analogues'', exhibited significantly greater ionisation parameters and sSFRs than the local, main-sequence, star-forming galaxies. Both \cite{2014MNRAS.442..900N} and \cite{2016ApJ...816...23S} show that the $\ott$ ratio, a strong tracer of the ionisation parameter, has a strong positive correlation with sSFR, a weak negative correlation with \Mstar\ and no clear correlation with SFR (as in Figures \ref{fig:qz_M} to \ref{fig:qz_MandSFR}). \cite{2015ApJ...812L..20K} find a strong positive correlation between the ionisation parameter and equivalent width of \Hb\ (EW(\Hb)). As discussed in \cite{2012MNRAS.419.1402G} and \cite{2015ApJ...812L..20K}, the EW(\Hb) traces the age of the stellar population, where the value of EW(\Hb) increases with the relative proportion of young to old stars. The sSFR we measure (based on the \Ha\ luminosity) reflects the number of massive young (OB) stars formed in relation to the total number of stars and is therefore sensitive to the fraction of young to old stars, similar to the EW(\Hb) and EW(\Ha).

        Based on our results, we argue that there exists an evolution in the ionisation parameter with redshift that is driven by the evolving sSFR. The sensitivity of the ionisation parameter to the sSFR may ensue from a relative increase in the hydrogen ionising photon flux with respect to the gas density with greater sSFR. Because the hydrogen ionising photons stem from massive young stars, the ``effective'' ionisation parameter of a galaxy will increase along with the relative proportion of young to old stars, or, the number of young stars per unit volume. Thus, the ionisation parameter would be expected to correlate strongly with sSFR (and SFR density). An increase in SFR alone is not sufficient to result in an increased ionisation parameter (Figures \ref{fig:qz_SFR} and \ref{fig:qz_global}) because a higher SFR does not necessarily increase the mean hydrogen ionising photon flux across an entire galaxy. Low-z galaxies with equivalent SFRs typically have higher \Mstar\ and lower sSFR (Fig. \ref{fig:match_global}) than galaxies at $z\geq 1.5$, indicating a lower fraction of young to old stars and/or lower SFR densities. 

        We note that our interpretation differs from some of the previous high-z studies we have compared our work to \citep[e.g][]{2014MNRAS.442..900N,0004-637X-801-2-88,2016ApJ...816...23S}. Based on the results presented here, we argue that high-z, main-sequence galaxies exhibit higher ionisation parameters than low-z, main-sequence galaxies. We attribute this difference in ionisation parameters, and the observed difference in positions on some emission-line ratio diagrams, to the evolution of the sSFR. There are three main reasons why we attribute the observed evolution of the ionisation parameter to the sSFR, rather than metallicity or \Mstar. Firstly, we use the metallicity as a second axis in our work, and thereby show clearly that even in cases where the metallicity of samples may be similar the ionisation parameters are clearly offset. Secondly, we directly compare matched low- and high-z samples, allowing us to disentangle the importance of \Mstar\ and sSFR. Thirdly, we infer the metallicity and ionisation parameter using combinations of emission-line ratios rather than using a single line ratio as a proxy, which could lead to misinterpretations regarding which ISM property is varying. 

        Although we have focussed on the metallicity and ionisation parameter, we also determine the electron density of our low-z matched samples via the $\SII\lambda6717/\SII\lambda6731$ ratio. We find no clear correlation between the electron densities and ionisation parameters of our samples. Samples with the highest inferred electron densities (and ISM pressures) do not exhibit the highest ionisation parameters and vice versa. Based on the mean (and median) $\SII\lambda6717/\SII\lambda6731$ ratios, the typical electron densities of the \Mstar-, SFR- and sSFR-matched samples are $\sim 40, 120$ and $60\cmcub$ respectively. Although the SFR-matched sample has the highest inferred electron density, the ionisation parameters of the two SFR bins are $0.3-0.4\dex$ lower than for the two bins of sSFR in the sSFR-matched sample. Moreover, the ionisation parameters of the two \Mstar\ bins of the low-z, \Mstar-matched sample are only $<0.1\dex$ lower than for the high SFR-matched sample, despite significantly higher electron densities. These results indicate that the ionisation parameter does not scale directly with the electron density, and, cannot be used as a proxy to derive the electron density, as it was in \citep{2014ApJ...787..120S}.

        We showed in \cite{2017MNRAS.465.3220K}, that samples of low- and high-z galaxies, matched in SFR, have similar electron densities. In contrast, samples matched in SFR alone do not exhibit the same ionisation parameters. As discussed in \cite[][ and others]{2014ApJ...787..120S,2014MNRAS.442..900N,2016ApJ...816...23S} one can derive a simple scaling relation, 
        \begin{align}
            q^3 \propto Q_0 n_e \epsilon^2 \, ,
            \label{eq:fill_fac}
        \end{align}
        between the ionising photon production rate, $Q_0$, the electron density and the volume filling factor, $\epsilon$, assuming the case of a Str{\"o}mgren sphere in ionisation equilibrium. The ionising photon production rate, $Q_0$, scales directly with the SFR, for the \HII\ region considered, and thus equation \eqref{eq:fill_fac} implies that for a fixed value of SFR and electron density the ionisation parameter scales as $q \propto \epsilon^{2/3} $. If this simple scaling relation extends to star-forming galaxies then our results may imply that high sSFR galaxies have higher filling factors and therefore higher ionisation parameters. However, volume filling factors have yet to be measured with sufficient accuracy either locally or at high redshift. 



\section{Summary} 
    \label{sec:conclusion}

    We have investigated the evolution of the ionisation parameter by comparing a set of main-sequence, star-forming galaxies at $z\sim 1.5$, derived from the FMOS-COSMOS and COSMOS-\OII\ Surveys, to samples of star-forming galaxies at $z<0.3$, taken from SDSS. To separate the effects of \Mstar, SFR and sSFR on the evolution of the ionisation parameter we selected four ``matched'' comparison samples at $z<0.3$. The first sample contains galaxies with equivalent \Mstar\ to the galaxies in our $z\sim 1.5$ sample and is representative of main-sequence galaxies at $z\sim 0.1$, the majority of which have significantly lower SFRs than the $z\sim 1.5$ sample. The second sample is comprised of galaxies with the equivalent SFRs to the galaxies in our $z\sim 1.5$ sample and is biased towards higher \Mstar\ than the $z\sim 1.5$ sample. The third local comparison sample is matched in sSFR, and has lower stellar masses and SFRs than our high-z sample. The final local comparison sample contains galaxies with similar \Mstar\ and SFR to the galaxies in our high-z sample. 

    In order to derive properties that are representative of the $z\sim 1.5$ sample, we rely on a stacked analysis of the matched samples. We bin each sample according to the property in which it is matched, such that the distribution of the properties being matched is consistent for each pair of $z<0.3$ and $z\sim 1.5$ bins. We use the relative emission-line fluxes of the stacked spectra to diagnose the metallicity and ionisation parameter via two diagnostic methods; the bayesian inference algorithm \texttt{IZI} and the widely used KK04 diagnostic. We take into account the variation in ISM pressure of the samples being used when applying \texttt{IZI} by selecting photoionisation models with ISM pressures inferred from the measured $\SII\lambda6717/\SII\lambda6731$ and $\OII\lambda3729/\OII\lambda3726$ ratios. Although the values of metallicity and ionisation parameter are offset between the two diagnostic methods, the conclusions drawn from the two are equivalent.

    Our main conclusions are as follows:
    \begin{enumerate}[leftmargin=*]
        \item The ionisation parameter of main-sequence galaxies evolves by $0.4 \dex$ from $z\sim 0.1$ to $z\sim 1.5$. 
        \item The ionisation parameter evolves with the sSFR of star-forming galaxies. There is no evolution in the ionisation parameter from $z\sim 0.1$ to $z\sim 1.5$ when the change in sSFR is taken into account. 
        \item The evolution of the ionisation parameter is not the byproduct of the evolution of the metallicity. The ionisation parameters of galaxies at the same redshift vary by $\sim 0.5\dex$ for constant metallicity. Moreover, the ionisation parameter scales most strongly with sSFR whereas the metallicity increases with \Mstar. 
    \end{enumerate}

    By matching samples we have shown that star-forming galaxies with equivalently high sSFRs have similar ionisation parameters, regardless of the consistency of \Mstar, SFR or metallicity. We put forward a simple physical explanation for the increase in ionisation parameter with sSFR. The ionisation parameter is defined as the ratio between the hydrogen ionising photon flux and the number density of hydrogen atoms. A high ionisation parameter may therefore result from a high relative proportion of young to old stars and/or a greater number of young stars per unit volume, both of which are observed as a high sSFR.  

    Much observational work remains to be undertaken to understand the effects of the ionisation parameter and ISM pressure on the evolution of star-forming galaxies.  With the launch of JWST it will become possible to greatly increase the number of galaxies with emission-line spectroscopy, allowing for more detailed, statistical studies of the metallicity, ionisation parameter and ISM pressure at high redshift.  However, contributions from diffuse gas emission can have a significant impact upon the ISM properties derived from strong emission lines. Spatially resolved studies of star-forming galaxies, across different cosmic epochs, (for e.g. via integral field spectroscopy surveys) will help address this issue, yielding further insight into the evolution of metallicity, ISM pressure and the ionisation parameter as well as the effects of applying \ion{H}{II} region diagnostics to entire galaxies.  



\section{Acknowldgements} 
    \label{sec:acknowldgements}

    LK gratefully acknowledges support from an ARC Laureate Fellowship (FL150100113). B.G. gratefully acknowledges the support of the Australian Research Council as the recipient of a Future Fellowship (FT140101202). Parts of this research were conducted by the Australian Research Council Centre of Excellence for All Sky Astrophysics in 3 Dimensions (ASTRO 3D), through project number CE170100013. We gratefully acknowledge the contribution of H. Jabran Zahid to the collection of the FMOS data. We hereby thank the anonymous referee for the insightful comments which greatly improved this paper.

    This paper is based on data collected at the Subaru Telescope, which is operated by the National Astronomical Observatory of Japan as well as data obtained at the W.M. Keck Observatory, which is operated as a scientific partnership among the California Institute of Technology, the University of California and the National Aeronautics and Space Administration. We wish to recognise and acknowledge the very significant cultural role and reverence that the summit of Mauna Kea has always had within the indigenous Hawaiian community and acknowledge that we are fortunate to have the opportunity to conduct observations from this mountain. We also thank the MPA/JHU team for making their catalog of SDSS data public.



\bibliographystyle{mnras}
\bibliography{Kaasinen_ionisation_source}

\end{document}